\documentclass[a4paper,fleqn,usenatbib]{mnras}
\usepackage{graphicx}
\usepackage{amsmath}
\usepackage{amssymb}
\usepackage{comment}
\usepackage{subfigure} 

\title[Reversing cooling flows with AGN jets]{Reversing cooling flows with AGN jets: shock waves, rarefaction waves, and trailing outflows}
\author[F. Guo et al.]{Fulai Guo$^{1,2}$\thanks{E-mail: fulai@shao.ac.cn}, Xiaodong Duan$^{1,2}$ and Ye-Fei Yuan$^{3}$\\
$^{1}$Key Laboratory for Research in Galaxies and Cosmology, Shanghai Astronomical Observatory, Chinese Academy of Sciences, \\
80 Nandan Road, Shanghai 200030, China\\
$^{2}$University of Chinese Academy of Sciences, 19A Yuquan Road, 100049, Beijing, China\\
$^{3}$Key Laboratory for Research in Galaxies and Cosmology, Department of Astronomy, University of Science and Technology of China, \\
Hefei, Anhui 230026, China}

\begin{document}
\bibliographystyle{mnras} 

\pagerange{000--000} \pubyear{0000}
\maketitle

\label{firstpage}

\begin{abstract}

The cooling flow problem is one of the central problems in galaxy clusters, and active galactic nucleus (AGN) feedback is considered to play a key role in offsetting cooling. However, {\it how} AGN jets heat and suppress cooling flows remains highly debated. Using an idealized simulation of a cool-core cluster, we study the development of central cooling catastrophe and how a subsequent powerful AGN jet event averts cooling flows, with a focus on complex gasdynamical processes involved. We find that the jet drives a bow shock, which reverses cooling inflows and overheats inner cool core regions. The shocked gas moves outward in a rarefaction wave, which rarefies the dense core and adiabatically transports a significant fraction of heated energy to outer regions. As the rarefaction wave propagates away, inflows resume in the cluster core, but a trailing outflow is uplifted by the AGN bubble, preventing gas accumulation and catastrophic cooling in central regions. Inflows and trailing outflows constitute meridional circulations in the cluster core. At later times, trailing outflows fall back to the cluster centre, triggering central cooling catastrophe and potentially a new generation of AGN feedback. We thus envisage a picture of cool cluster cores going through cycles of cooling-induced contraction and AGN-induced expansion. This picture naturally predicts an anti-correlation between the gas fraction (or X-ray luminosity) of cool cores and the central gas entropy, which may be tested by X-ray observations. 

\end{abstract}


\section{Introduction}
\label{section:intro}

Presumably located at knots of the cosmic web, galaxy clusters are the largest gravitationally bound systems in the Universe. The baryon component of galaxy clusters is dominated by the hot, diffuse intracluster medium (ICM) at keV temperatures, which emits copiously in X-rays. One of the central problems in galaxy clusters is the so-called cooling flow problem, which states that gas inflows resulting from radiative cooling with typical mass inflow rates of hundreds to a thousand solar masses per year are inconsistent with current multi-wavelength observations (see \citealt{peterson06} for a review). It is now widely accepted that one or more heating sources are present in galaxy clusters, offsetting radiative cooling and suppressing the development of cooling flows.

One of the key physical mechanisms towards solving the cooling flow problem is radio-mode AGN feedback, which operates as radio-emitting jets apparently emanating from central supermassive black holes (SMBHs) penetrate through and heat the ICM (\citealt{mcnamara12}; \citealt{fabian12}; \citealt{heckman14}; \citealt{soker16}). The jet-ICM interaction inflates kpc-sized  active galactic nucleus bubbles (AGN bubbles) of relativistic plasma, which have been frequently observed through radio synchrotron emissions and in X-ray images of galaxy clusters as surface brightness depressions, i.e., ``X-ray cavities"  (e.g., \citealt{boehringer93}; \citealt{fabian02}; \citealt{birzan04}; \citealt{croston11}). Radio-mode AGN feedback also induces shocks (e.g., \citealt{randall11}; \citealt{randall15}) and sound waves (e.g., \citealt{fabian03b}; \citealt{fabian17}) in the ICM. Observations suggest that radio-mode AGN feedback is usually energetically sufficient to offset radiative cooling from cool cores of galaxy clusters (\citealt{birzan04}; \citealt{rafferty06}).

The key question then is if radio-mode AGN feedback can effectively suppress cooling flows during at least a major fraction of the lifetimes of galaxy clusters, which has been extensively investigated with hydrodynamic simulations (e.g., \citealt{ruszkowski02}; \citealt{brighenti06}; \citealt{vernaleo06}; \citealt{cattaneo07}; \citealt{guo08a}; \citealt{dubois11}). Recent simulations which implement momentum-driven AGN jets or outflows self-consistently triggered by cold gas accretion have achieved great success in averting catastrophic cooling while simultaneously keeping observed cool-core profiles in a time-averaged sense (e.g., \citealt{gaspari11}; \citealt{gaspari12}; \citealt{li14}; \citealt{li15}; \citealt{prasad15}; \citealt{yang16b}). However, to heat the ICM isotropically, the jets in these simulations are often assumed to be precessing with a typical precession angle of $\sim 10^{\circ}$-$25^{\circ}$. It is unclear if this assumption represents reality, although observational signatures of jet reorientation have been found in some clusters \citep{babul13}. Furthermore, AGN jets in clusters are presumably powered by SMBH's hot accretion flows \citep{yuan14}, which have never been directly simulated in cluster simulations due to the extremely large dynamical range involved (see relevant discussions in \citealt{gaspari17}). Therefore, there is no guarantee that the sub-grid treatment of SMBH accretion, a key part of the AGN feedback loop, is correct in current simulations.

A related central question is {\it how} AGN jets deliver energy to the ICM and avert cooling flows, which has been actively debated in literature. Proposed channels of AGN heating include the pdV work of expanding AGN bubbles (\citealt{ruszkowski02}; \citealt{guo08b}), shock heating (\citealt{bruggen07}; \citealt{randall15}; \citealt{liyuan17}), viscous dissipation of sound waves (\citealt{ruszkowski04a}; \citealt{fabian17}), mixing of the ICM with jet plasma (\citealt{hillel16}; \citealt{hillel17}), turbulence dissipation (\citealt{zhuravleva14}; \citealt{zhuravleva16}), and cosmic ray heating through the dissipation of self-triggered hydromagnetic waves (\citealt{guo08a}; \citealt{jacob17}; \citealt{ruszkowski17}). The pdV work does not directly increase the ICM's entropy, but it increases the ICM's thermal energy, which is subsequently converted to its kinetic and potential energies during the ICM motions and may eventually dissipate into heat and entropy. It is likely that all of these heating channels take place to some extent in jet-mode AGN feedback, but it has been not easy to perform a clean analysis on their relative importance, which may also depend on jet properties \citep{tang17}.

In this paper, we use a specific idealized simulation of a typical cool-core cluster to investigate in detail how a powerful AGN jet event triggered by the central cooling catastrophe averts and even reverse cooling flows. Unlike many recent simulations focusing on the jet's heating effects and the ICM's long-term balance between heating and cooling, we particularly focus on complex gasdynamical processes during and after the AGN feedback event, including the bow (forward) shock, rarefaction wave, trailing outflow uplifted by the AGN bubble, meridional circulations in the cluster core. We investigate their dynamical and energetic roles in heating and reversing cooling inflows and in averting catastrophic cooling. Such a detailed analysis on a specific AGN jet event would be complementary to other studies that aim to investigate the long-term ability of AGN feedback in solving the cooling flow problem.

The rest of the paper is structured as follows. We describe our model and the numerical setup in Section 2, and present our results in Section 3. In Section 3.1, we study the overall evolution of the ICM's temperature, density and entropy profiles before and after the AGN jet event. We then analyze the bow shock and rarefaction wave induced by the AGN event in Section 3.2. We study how the AGN event heats the ICM in our simulation in Section 3.3, and the role of the rarefaction wave in energy redistribution in the ICM in Section 3.4. We then study trailing outflows uplifted by the AGN bubble and cool-core circulations along meridional planes in Section 3.5, and describe an emerging picture of cool cluster cores going through cycles of cooling-induced contraction and AGN-induced expansion in Section 3.6. We finally summarize and discuss our results in Section 4.

\section{Methodology}
\label{section2}

\subsection{Equations and Numerical Setup}

We study the thermal and hydrodynamic evolution of the ICM gas under the influence of gravity, radiative cooling, and AGN feedback. The basic hydrodynamic equations governing the ICM evolution may be written as 
\begin{eqnarray}
\frac{d \rho}{d t} + \rho \nabla \cdot {\bf v} = 0,\label{hydro1}
\end{eqnarray}
\begin{eqnarray}
\rho \frac{d {\bf v}}{d t} = -\nabla P-\rho \nabla \Phi,\label{hydro2}
\end{eqnarray}
\begin{eqnarray}
\frac{\partial e}{\partial t} +\nabla \cdot(e{\bf v})=-P\nabla \cdot {\bf v} - \mathcal{C}
   \rm{ ,}\label{hydro3}
   \end{eqnarray}
  \\ \nonumber
\noindent
where $d/dt \equiv \partial/\partial t+{\bf v} \cdot \nabla $ is the Lagrangian time derivative, $\Phi$ is the gravitational potential, and $\rho$, ${\bf v} $, $e$, $P$ are the density, velocity, energy density, pressure of the ICM gas respectively. $\mathcal{C}$ is the energy loss rate per unit volume due to radiative cooling. The above hydrodynamic equation set is closed by the relation $P=(\gamma-1)e$, where $\gamma=5/3$. 

In this paper, we specifically study the evolution of the ICM in a typical cool-core galaxy cluster subject to cooling and an AGN jet episode. The most basic picture here contains a rotational symmetry with respect to the jet axis. We thus solve Equations (\ref{hydro1}) - (\ref{hydro3}) in $(R, z)$ cylindrical coordinates using our own two-dimensional Eulerian code similar to ZEUS 2D \citep{stone92}. The code has been extensively used in many previous studies, e.g., \citet{guo10b}, \citet{guo10a}, \citet{guo11}, \citet{guo12}, \citet{guo12b}, \citet{guo15}, \citet{guo16}. Starting from the origin, the computational grid consists of $800$ equally spaced zones in both coordinates out to $200$ kpc plus additional $400$ logarithmically-spaced zones out to $2$ Mpc. The spatial resolution within the inner 200 kpc is 0.25 kpc, while the large outer boundary ($2$ Mpc) minimizes numerical impacts of the outer boundary conditions on the cool cluster core region. We adopt reflective boundary conditions at the inner boundaries. At the outer boundaries, we adopt reflective boundary conditions as well, which prohibit mass exchange between active and ghost zones \citep{stone92}, facilitating the analysis of the ICM energy evolution in Section 3.3.

The radiative cooling rate in Equation (\ref{hydro3}) may be written as $\mathcal{C}=n_{\rm i}n_{\rm e}\Lambda(T,Z)$, where $n_{\rm e}$ is the electron number density, $n_{\rm i}$ is the ion number density, and the cooling function $\Lambda(T,Z)$ is adopted from \citet{sd93} with a fixed metallicity $Z=0.4$. The cooling time often becomes very short when the gas cools to low temperatures. In order to accelerate the computations, we enforce a minimum gas temperature floor $T_{\rm floor}=10^{5}$ K, and employ the sub-cycling method for radiative cooling (\citealt{anninos97}; \citealt{proga03}; \citealt{ruszkowski17}) when the local cooling time becomes shorter than the hydrodynamical timestep. Our simulations do not have enough spatial resolutions and physical processes to correctly follow the evolution of cold gas.

In the collisionally-ionized hot ICM, the ion number density $n_{\rm i}$ may be roughly approximated to be $n_{\rm i}=1.1n_{\rm H}$, where $n_{\rm H}$ is the proton number density. Therefore the mean molecular weight per particle is $\mu=0.61$, and that per electron is $\mu_{\rm e}=5\mu/(2+\mu)=1.17$. According to the ideal gas law, the gas temperature is related to the gas pressure and density via:
\begin{eqnarray}
T=\frac{\mu m_{\mu}P}{k_{\rm B} \rho}= \frac{\mu P}{\mu_{\rm e} k_{\rm B} n_{\rm e}}{\rm ,}
   \end{eqnarray}
where $k_{\rm B}$ is Boltzmann's constant and $m_{\mu}$ is the atomic mass unit. 

\subsection{Cluster Setup}

 \begin{figure}
\includegraphics[width=0.45\textwidth]{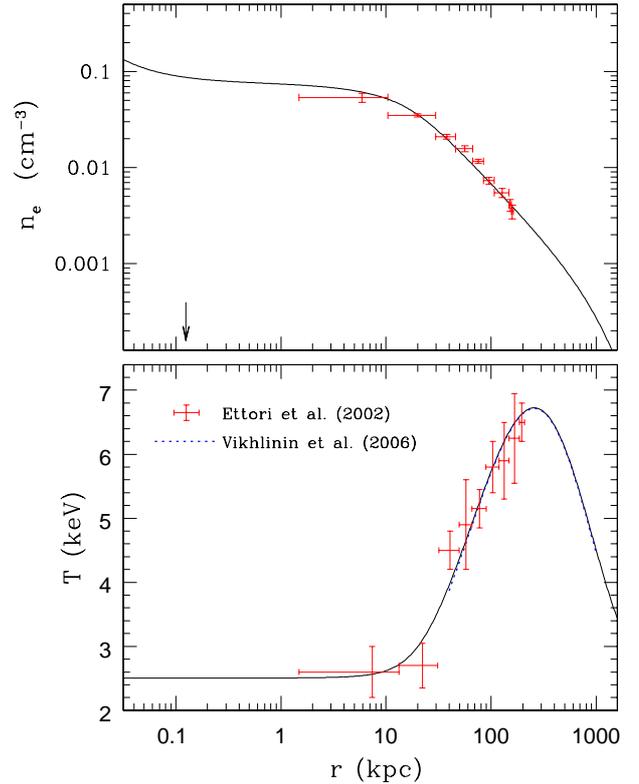} 
\caption{Initial density and temperature profiles of the ICM gas in our simulation (solid lines). Crosses indicate {\it Chandra} data of Abell 1795 by \citet{ettori02}, while the dotted line in the bottom panel represents the best analytical fit to {\it Chandra} data by \citet{vikhlinin06} covering between $40$ to $1000$ kpc. The arrow in the top panel indicates the centre of the innermost active zone along either the $R$ or $z$ direction in our simulation.}
 \label{plot1}
 \end{figure} 

For initial conditions, we construct an idealized cool core galaxy cluster in hydrostatic equilibrium. For concreteness, the cluster profiles are initialized according to the archetypal cool core cluster Abell 1795, which has been well observed by both {\it Chandra} and {\it XMM-Newton} \citep{tamura01, ettori02, vikhlinin06}. The initial cluster setup is very similar to that in \citet{guo10b} and \citet{guo14}, where the readers are referred to for more details. 

We first setup the radial temperature profile of the ICM using an analytic fit to the deprojected $3$-dimensional temperature distribution of A1795 derived from {\it Chandra} observations:
\begin{eqnarray}
T(r)=(T_{\rm in}^{5}+T_{\rm V}^{5})^{1/5}{,}
\end{eqnarray}
where $T_{\rm in}=2.5$ keV is the observed central temperature of A1795, and $T_{\rm V}$ is the best-fit temperature profile to {\it Chandra} data of A1795 proposed by \citet{vikhlinin06}:
\begin{eqnarray}
T_{\rm V}(r)=9.68t_{\rm cool}(r)t_{\rm out}(r)\text{ keV,}
\end{eqnarray}
where 
\begin{eqnarray}
t_{\rm cool}(r)=\frac{0.1+(r/r_{1})^{1.03}}{1+(r/r_{1})^{1.03}}\text{ ,}
\end{eqnarray}
describes that the temperature declines inward in the inner cool-core region, and
\begin{eqnarray}
t_{\rm out}(r)=\frac{1}{[1+(r/r_{2})^{a}]^{b/a}}\text{ ,}
\end{eqnarray}
represents the radially declining outer region. Here the parameters are $r_{1}=77$ kpc, $r_{2}=550$ kpc, $a=1.63$ and $b=0.9$ \citep{vikhlinin06}. Our initial gas temperature profile $T(r)$ is shown in the bottom panel of Figure \ref{plot1}, indicating clearly that it provides a reasonably good fit to {\it Chandra} X-ray data of A1795 covering from the inner few kpc to $\sim 1$ Mpc.

The initial gas density profile is derived from the assumption of hydrostatic equilibrium and a fixed gravitational potential contributed by three components: the dark matter halo ($\Phi_{\text{DM}}$), the central galaxy ($\Phi_ {*}$), and the central SMBH ($\Phi_{\text{BH}}$):
\begin{eqnarray}
\Phi = \Phi_{\text{DM}} +  \Phi_ {*} + \Phi_{\text{BH}}  \text{.}  
\end{eqnarray}
\noindent
We adopt a Navarro-Frenk-White (NFW) profile \citep{navarro97} for the dark matter density distribution:
\begin{eqnarray}
\rho_{\text{DM}}(r)=\frac{M_{0}/2\pi}{r(r+r_{\text{s}})^{2}}\text{,}  
\end{eqnarray}
\noindent
where $r_{\text{s}}$ is the standard scale radius and $M_{0}$ is a characteristic mass. As in \citet{guo14}, we take $M_{0}=4.2\times 10^{14}M_{\odot}$, and $r_{\text{s}}=430$ kpc for A1795, which correspond to a virial mass $M_{\rm vir}=7.06\times 10^{14}M_{\odot}$ and a viral radius $r_{\rm vir}=1.8$ Mpc at the redshift of A1795 $z=0.0632$. Here $r_{\rm vir}$ is defined as the radius within which the mean dark matter density equals to $200 \rho_{\rm crit}(z)$, where $\rho_{\rm crit}(z)=3H(z)^{2}/8\pi G$ is the critical density of the universe at the redshift of the system. The corresponding dark matter gravitational potential is:
\begin{eqnarray}
\Phi_{\text{DM}}= -\frac{2GM_{0}}{r_{\text{s}}} \frac{\text{ln}(1+r/r_{\text{s}})}{r/r_{\text{s}}}      \text{,}  
\end{eqnarray}
\noindent
where $G$ is the gravitational constant. 

We take the stellar mass density of the central galaxy to have a Hernquist profile \citep{hernquist90}:
\begin{eqnarray}
\rho_ {*} = \frac{M_ {*}a}{2\pi r}\frac{1}{(r+a)^{3}}     \text{,}  
\end{eqnarray}
\noindent
where $M_ {*}$ is the total stellar mass and $a=R_{\rm e}/1.8153$ is a scale length. Here $R_{\rm e}$ is the radius of the isophote enclosing half the galaxy's light. For the Brightest Cluster Galaxy (BCG) in A1795, $M_ {*}=5.35\times 10^{11}M_{\odot}$ and $R_{\rm e}=40.3$ kpc (\citealt{chandran07}; \citealt{guo14}). We note that the adopted value of $R_{\rm e}$ is larger than typical BCG sizes, which may underestimate the BCG's contribution to the gravitational potential:
\begin{eqnarray}
\Phi_{*}= -\frac{GM_ {*}}{r+a}     \text{.} 
\end{eqnarray}

The central SMBH's gravitational potential may be written as
\begin{eqnarray}
\Phi_{\rm BH}= -\frac{GM_{\rm BH}}{r-r_{\rm g}}     \text{,} 
\end{eqnarray}
\noindent
where $M_{\rm BH}$ is the SMBH's mass, $r_{\rm g}=2GM_{\rm BH}/c^{2}$ is the Schwarzschild radius, and the $1/(r-r_{\rm g})$ mimics the effects of general relativity (\citealt{paczynsky80}; \citealt{quataert00}). For the central SMBH in A1795, we take $M_{\rm BH}=1.66\times 10^{9}M_{\odot}$ \citep{chandran07}. While we include the gravitational acceleration contributed by the central SMBH in our simulations, it is negligibly small compared to that contributed by the dark matter halo and the central galaxy in all computational zones. 

The initial cluster profiles described here are spherically symmetric. We derive the initial gas density profile from the initial temperature profile and the gravitational potential, assuming hydrostatic equilibrium. We normalize the density profile by the central electron number density $n_{\rm 0}$ (more specifically, the value of $n_{\rm e}$ at $r=9.8$ pc), which is used as a free parameter to fit the observed radial gas density distribution. In our simulations, the value of $n_{0}$ is chosen to be $n_{0}=0.48$ cm$^{-3}$, which leads to an initial density profile that provides a reasonably good fit to observations, as shown in the top panel of Figure \ref{plot1}.

 \begin{figure*}
 \centerline{
\includegraphics[width=0.45\textwidth]{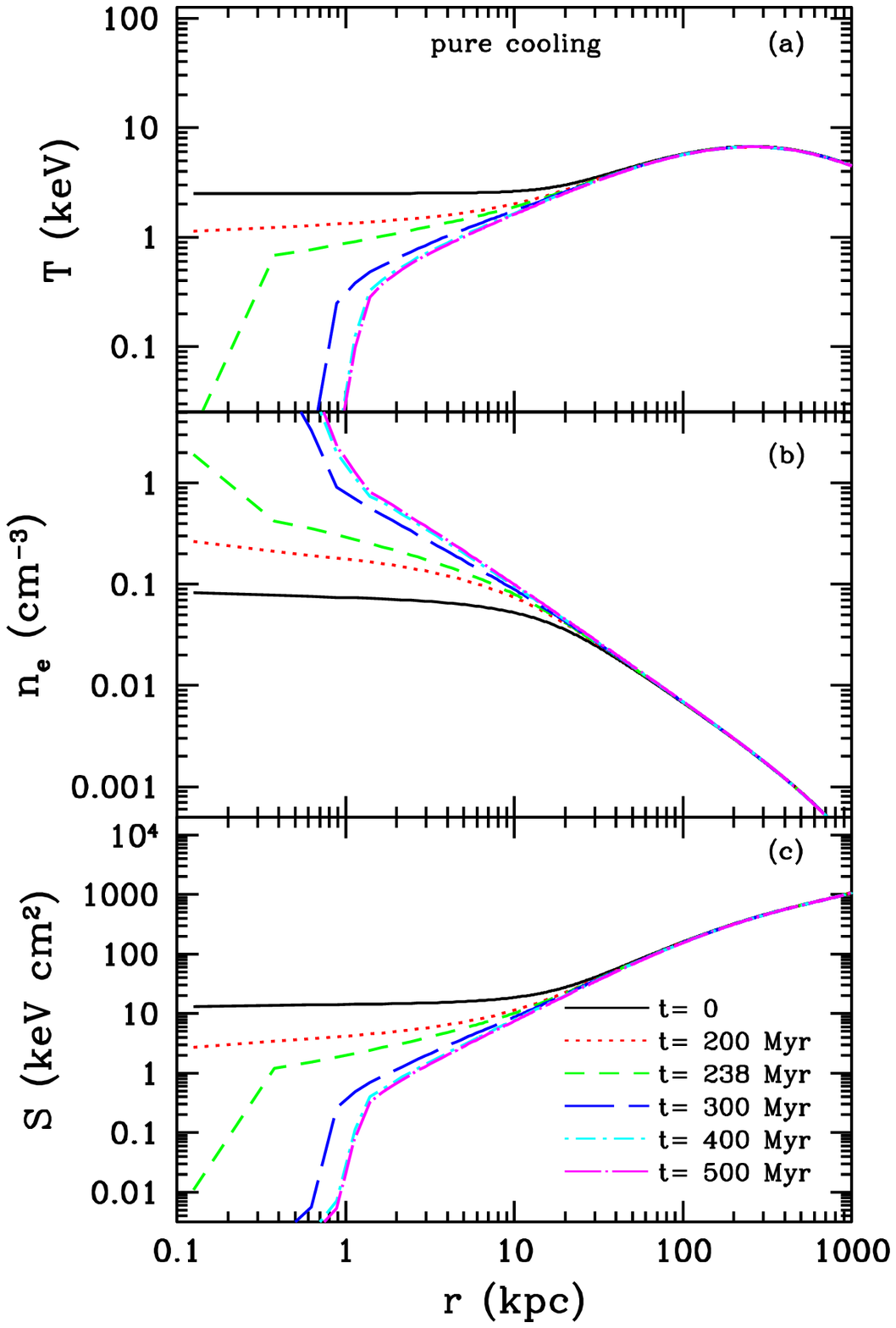} 
\includegraphics[width=0.45\textwidth]{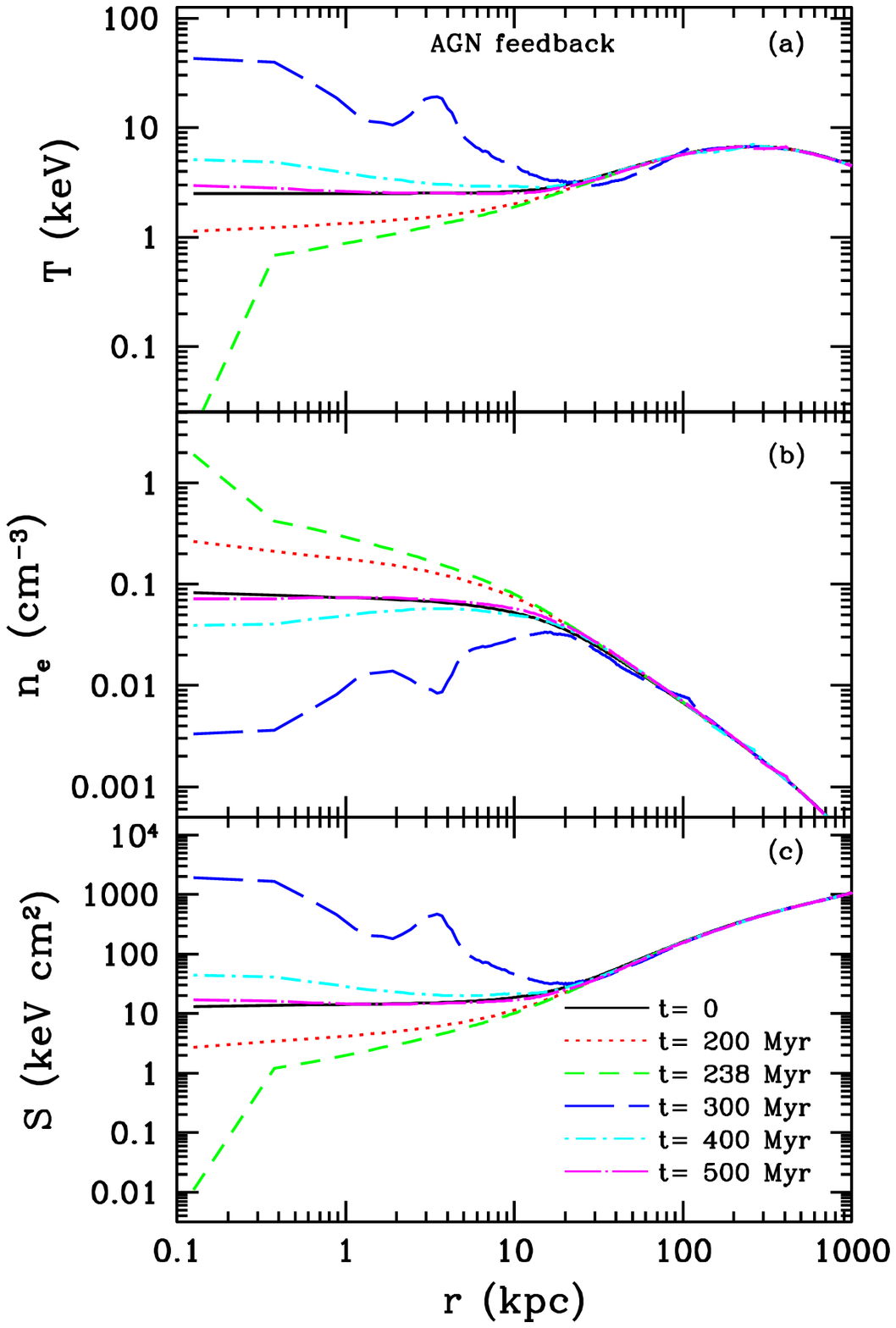} }
\caption{Temporal evolution of mass-weighted spherically-averaged profiles of (a) gas temperature (b) electron number density and (c) specific entropy $S\equiv T/n_{\rm e}^{2/3}$. The left panels correspond to the ``pure cooling" run, where no AGN jets are turned on during the whole simulation. In our main ``AGN feedback" run (right panels), the ICM first evolves due to gravity and radiative cooling until $t \sim 238$ Myr, when a central cooling catastrophe develops. Then an episode of AGN jet activity with a duration of $t_{\rm jet}=5$ Myr is triggered, and over-heats the cool cluster core. At $t\sim 500$ Myr, the ICM returns roughly back to its initial profile due to cooling.}
 \label{plot2}
 \end{figure*} 
 
\subsection{The Numerical Experiment and Jet Modeling}
\label{section:experiment}

With the initial setup described in the previous two subsections, we performed a numerical experiment to study the development of cooling flows and more importantly, how a subsequently-triggered powerful AGN outburst transfers energy to the ICM, averting cooling inflows. AGN heating in galaxy clusters is usually considered as a feedback mechanism, which relates the onset of AGN outbursts directly to the ICM cooling and inflow. Such an AGN feedback mechanism is required to stably maintain cool-core cluster profiles, as shown in the global stability analysis of \citet{guo08b}. Recent state-of-the-art hydrodynamical simulations in this field also favor the self-regulated AGN feedback mechanism, where AGN outbursts are triggered by the so-called ``cold-mode accretion" of cold gas dropped out from the hot ICM (\citealt{gaspari11}; \citealt{gaspari12}; \citealt{li15}; \citealt{yang16}; \citealt{yang16b}; \citealt{liyuan17}). These simulations usually start with a spherically-symmetric hydrostatic ICM, and cold gas often first appears right after the central cooling catastrophe happens. 

Our numerical experiment is in line with recent progress. We first let the initially hydrostatic cluster evolve under the influence of radiative cooling and gravity, but without AGN outbursts. Once the central cooling catastrophe happens (after around the central gas cooling time; \citealt{guo14}), AGN feedback is manually triggered. To be specific, we turn on an AGN jet in our simulated hemisphere of the cluster once the total mass of cold gas with $T<T_{\rm crit}=5\times 10^{5}$ K within the central $1$ kpc becomes larger than $M_{\rm crit}=10^{-3}M_{\odot}$. Once the cooling catastrophe starts, it develops very quickly, and thus the exact values of $T_{\rm crit}$ and $M_{\rm crit}$ do not affect the results in the paper. For simplicity, when it is active, the jet in our simulation is steady and uniform at the jet base, corresponding to a steady accretion flow around the central SMBH. This is different from the ``instantaneous" feedback model adopted in recent simulations (e.g., \citealt{yang16b}; \citealt{liyuan17}), where the jet parameters are determined instantaneously by the amount of central cold gas and a fixed ``free-fall" timescale. In reality, before being accreted by the central SMBH, cold gas may circularize around it, forming a torus or accretion flow. The jet properties are determined by complex physical processes taking place in accretion flows. Our simple jet modeling method provides a straightforward way to investigate how a powerful AGN jet episode suppresses and even reverses cooling inflows.

In our simulation, the AGN jet is injected into the ICM at the cluster centre along the $z$-axis with an opening angle of $0$ degree. Similar to \citet{brueggen07} and \citet{guo11}, we initialize the jet by applying inflow boundary conditions to a cylindrical nozzle placed in ghost zones, which inject the mass-, momentum-, and thermal energy fluxes into active computational zones. The jet's kinetic energy is automatically injected into the ICM with the injection of mass and momentum. In the initialization nozzle, the jet is uniform and steady, characterized by $4$ parameters: density $\rho_{\rm jet}$, thermal energy density $e_{\rm jet}$, velocity $v_{\rm jet}$, and radius $R_{\rm jet}$. The jet is turned off after an active duration of $t_{\rm jet}$. In our simulation, the values of these parameters are: $\rho_{\rm jet}=1.61\times 10^{-26}$ g cm$^{-3}$ (density contrast $\eta =0.1$ with respect to the central ICM density in the innermost active zone at $t=0$), $e_{\rm jet}=1.91\times 10^{-9}$ erg cm$^{-3}$ (energy density contrast $\eta_{\rm e} \equiv e_{\rm jet}/e_{\rm amb}=2$ with respect to the central ICM thermal energy density $e_{amb}$ at $t=0$), $v_{\rm jet}=3.0\times 10^{9}$ cm s$^{-1}$, $R_{\rm jet}=1.5$ kpc ($6$ computational grids along the $R$ direction), and $t_{\rm jet}=5$ Myr. These values are specifically chosen to ensure that the AGN event is powerful enough to reheat the cooling flow back to the initial cluster profile, while driving a bow shock with its aspect ratio (major to minor axis ratio) roughly consistent with observations. From $t=250$ Myr to $300$ Myr as seen in the left panel of Figure \ref{plot3}, the shock's aspect ratio evolves from $3.0$ to $1.6$. For comparison, the shock's aspect ratios in two observed powerful AGN outbursts Cygnus A \citep{wilson06} and MS 0735.6+7421 (hereafter MS0735; \citealt{gitti07}) are $2.2$ and $1.5$, respectively (considering the projection effect, the real values may be larger; \citealt{mathews10}). According to the classification of AGN jets in \citet{guo15} and \citet{guo16}, the jet in our simulation is light, and internally supersonic with an internal Mach number $M_{\rm int} \sim 8.26$.

The jet power in the initialization nozzle may be written as
\begin{eqnarray}
P_{\rm jet}=P_{\rm kin} +P_{\rm th}{\rm ,}
 \end{eqnarray}
where $P_{\rm kin}=\rho_{\rm jet} v_{\rm jet}^{3}\pi R_{\rm jet}^{2}/2$ and $P_{\rm th}=e_{ \rm jet} v_{\rm jet}\pi R_{\rm jet}^{2}$ are the kinetic and thermal powers of the jet, respectively. The jet in our simulation is dominated by the kinetic power with $P_{\rm kin}=1.46\times10^{46}$ erg/s and $P_{\rm th}=3.85\times 10^{44}$ erg/s. The total jet power $P_{\rm jet}=1.5\times10^{46}$ erg/s is close to that estimated in Cygnus A \citep{wilson06} and MS0735 \citep{mcnamara09}, but less than the maximum jet power $\sim 10^{47}$ erg/s reached in some recent simulations (e.g., \citealt{gaspari11}; \citealt{li15}; \citealt{yang16b}). The total jet energy in our simulation is $E_{\rm jet}=P_{\rm jet}t_{\rm jet}\sim 2.4\times 10^{60}$ erg, which is much less than that in Cygnus A and MS0735. The jet episode in our simulation is powerful and brief, while a less powerful jet with a longer duration would result in a more gentle impact on cluster cores. The dominant channel through which the jet delivers energy to the ICM may be different for these two types of jet activities.

\section{Results}
\label{section:results}

In this section, we present the results of our main simulation (thereafter denoted as the ``AGN feedback" run) described in detail in Section \ref{section:experiment}. For comparison, we also present some results from a control run, denoted as the ``pure cooling" run, where cooling flows are developed unimpeded without turning on AGN feedback throughout the whole run. 

\subsection{Evolution of ICM Profiles}

We first investigate the overall cluster evolution by looking at the spherically-averaged profiles. Figure \ref{plot2} shows the time evolution of mass-weighted spherically-averaged profiles of gas temperature, density, and specific entropy $S\equiv T/n_{\rm e}^{2/3}$ in the ``pure cooling" run (left panels) and our main ``AGN feedback" run (right panels). The left panels clearly show the development of cooling flows under the influence of gravity and radiative cooling. The hot ICM initially in hydrostatic equilibrium gradually loses pressure support due to cooling, leading to a gradual decrease in both temperature and entropy, and increase in density. At $t \sim 238$ Myr, a central cooling catastrophe suddenly appears, and the gas temperature in central regions drops very quickly down to the temperature floor in our simulation. After that, cold gas drops out and accumulates in central regions quickly. The mass inflow rate across $r=50$ kpc in this ``catastrophic cooling" state of A1795 is about $160$ - $200 M_{\odot}/$yr, as seen in Figure \ref{plot9} and relevant discussions in Section 3.6. This picture on the cooling-flow development in galaxy clusters is consistent with recent high-resolution simulations in \citet{li12} and \citet{guo14}.

The cluster evolution in our main run is shown in the right panels. During the first $238$ Myr,  cooling flows develop gradually in the ICM without AGN feedback, which is the same as in the ``pure cooling" run. Once the central cooling catastrophe appears at $t \sim 238$ Myr, an AGN jet activity is manually turned on to mimic AGN feedback triggered by the SMBH's accretion of cold gas resulting from the central cooling catastrophe. As elaborated in Section \ref{section:experiment}, the jet is specifically set up to be powerful enough to avert cooling flows, while its power is still consistent with powerful AGN outbursts observed in some cool core clusters (e.g., Cygnus A and MS0732).

The right panels of Figure \ref{plot2} clearly show that the jet stops the central cooling catastrophe, and heats up the cluster core. In the inner regions with $r\lesssim 20$ kpc, the gas is over-heated as the temperature and entropy rise above their initial values at $t=0$. The jet event in our simulation is brief with a duration of $t_{\rm jet}=5$ Myr. After the jet event, the cluster re-adjusts its profiles, and cools later on back to roughly its initial distribution at $t\sim 500$ Myr. In the subsections below, we focus on studying in detail the energetical and dynamical processes as the jet averts and even reverses the cooling inflow.

 \begin{figure*}
 \centerline{
\includegraphics[height=0.8\textheight]{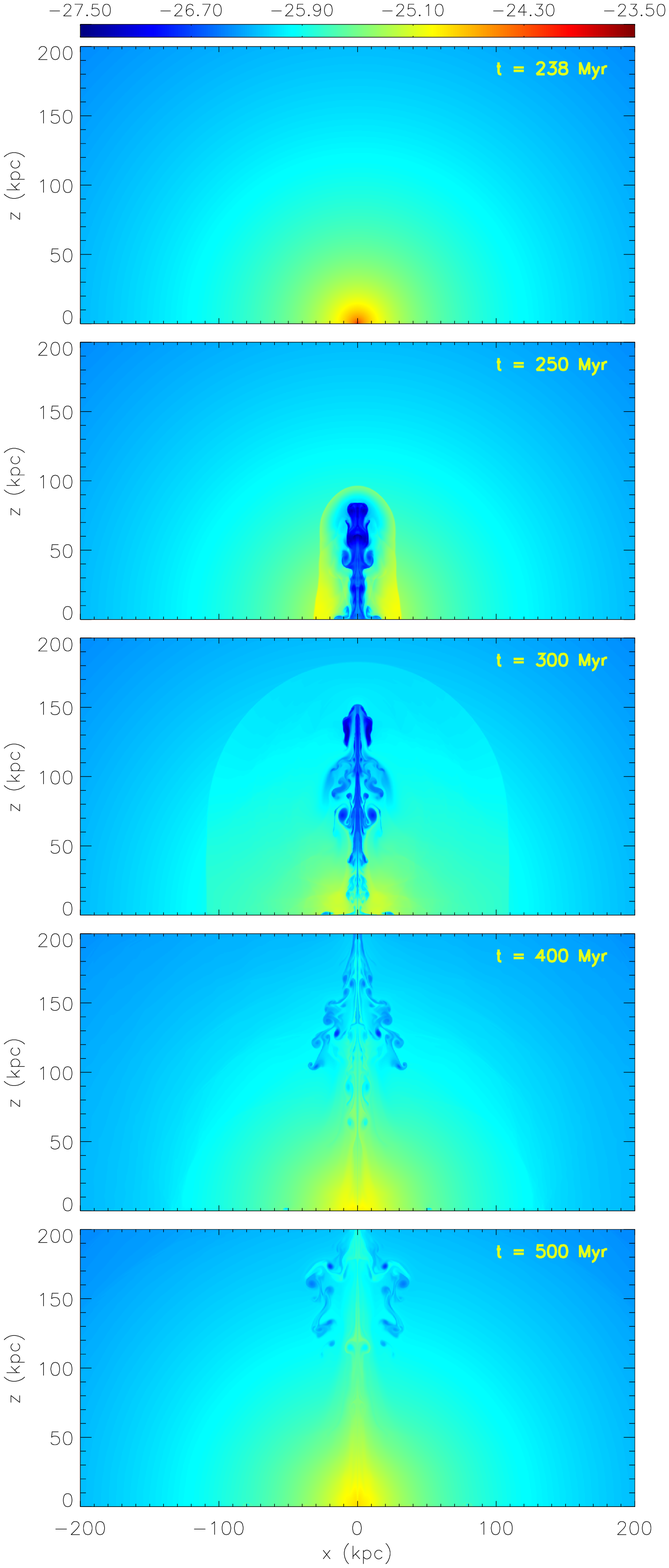} 
\includegraphics[height=0.8\textheight]{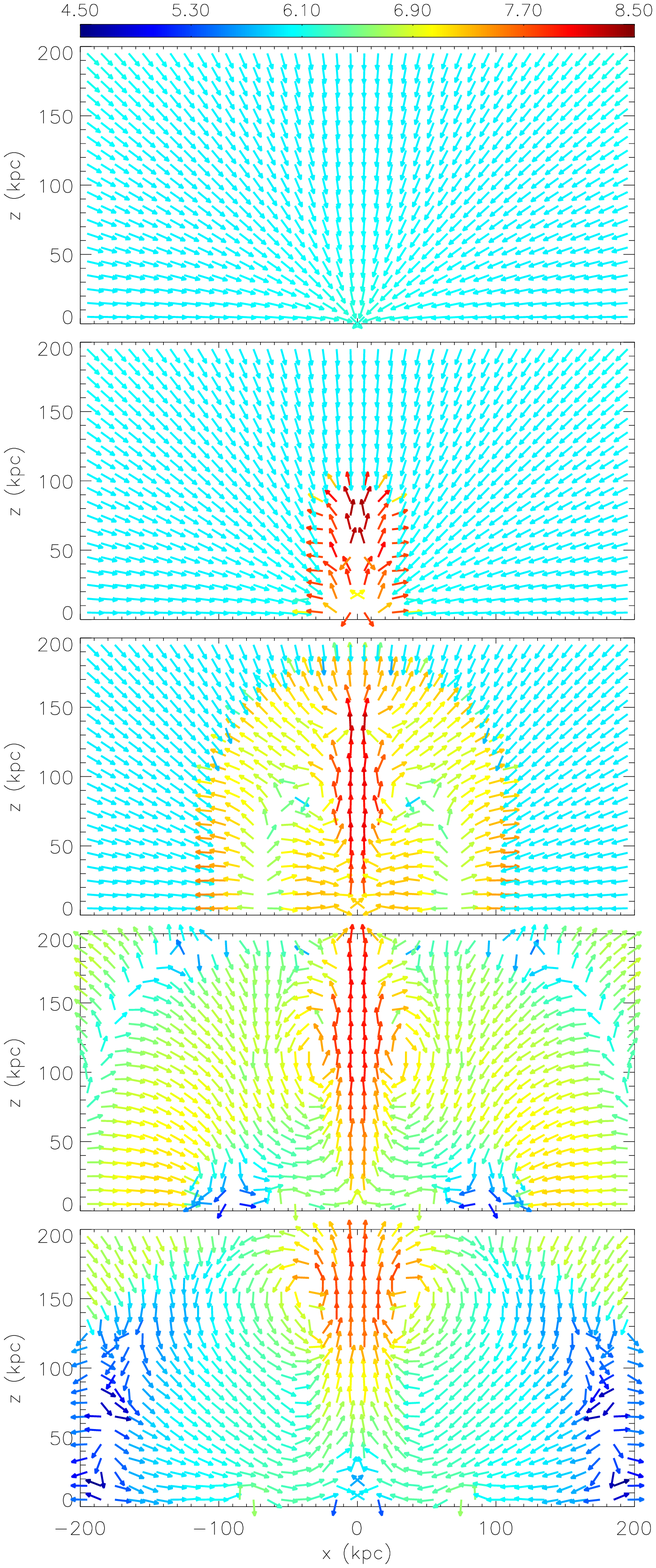} }
\caption{Temporal evolution of the inner cluster region ($400 \times 200$ kpc) in our main ``AGN feedback" run. The left panels show the gas density distribution at five times $t=238$, $250$, $300$, $400$, and $500$ Myr, and the top colour bar refers to gas density in units of g cm$^{-3}$ in logarithmic scale. The corresponding velocity distributions at these five times are shown in the right panels, where arrows indicate velocity directions, while their colours refer to velocity magnitudes with a colour bar in units of cm/s in logarithmic scale shown above the right panels. The jet is turned on at $t=238$ Myr, right after the time of the top panels.}
 \label{plot3}
 \end{figure*}
 
 \begin{figure}
\includegraphics[width=0.45\textwidth]{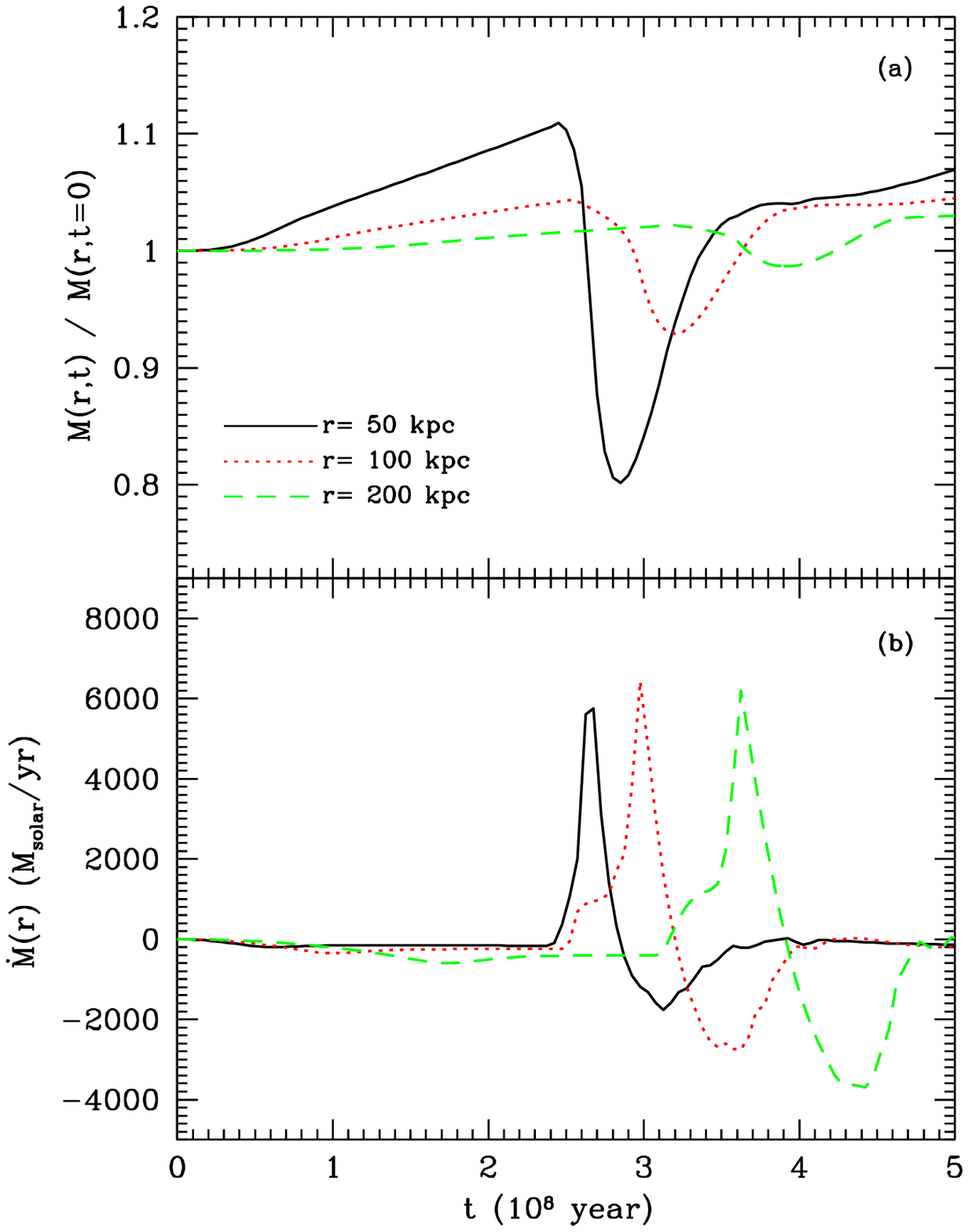} 
\caption{{\it Top panel}: Temporal evolution of integrated gas masses within three spheres of radii $r= 50$ ({\it solid}), $100$ ({\it dotted}), and $200$ kpc ({\it dashed}) normalized by the corresponding enclosed gas masses at $t=0$. Durations with positive slopes indicate ICM inflows and cluster contraction, while negative slopes indicate outflows and expansion. {\it Bottom panel}: Time evolution of gas mass flow rates (averaged with a moving $5$ Myr window) across the surfaces of three spheres with radii $r= 50$ ({\it solid}), $100$ ({\it dotted}), and $200$ kpc ({\it dashed}). Positive values indicate outflows, while negative values correspond to inflows.}
 \label{plot4}
 \end{figure} 
 
\subsection{Shock Waves and Cool Core Expansion}

What happens to the cooling inflow as the jet is injected into the ICM? Figure \ref{plot3} shows the temporal evolution of the inner cluster region in our simulation. The left and right panels show the gas density and velocity distributions, respectively, at $t=237.6$, $250$, $300$, $400$, and $500$ Myr (from top to bottom). The top panels correspond to the ICM state at $t=237.6$ Myr, right before the onset of the central cooling catastrophe. Clearly the ICM gas in the cooling flow is all moving toward the cluster centre, where the gas density is the highest and the cooling time is the shortest. Which naturally explains why the cooling catastrophe first appears in the cluster centre.

The evolution of AGN jets in an ambient medium has been extensively studied with hydrodynamic simulations in the past decades (e.g., \citealt{norman82}; \citealt{reynolds01}; \citealt{ruszkowski04a}; \citealt{hardcastle13}; \citealt{guo15}). As a supersonic jet penetrates through the ambient ICM, it drives a bow shock that encloses the jet and the lobe resulting from its interaction with the ICM. Since the jet moves much faster than the bow shock, it creates a hotspot at its head (working surface) bracketed by the apex of the bow shock and a reverse shock within the jet near its working surface. The jet's kinetic energy is quickly dissipated into thermal energy through the reverse shock, and the high pressure in the hot spot drives expansion and the bow shock. The outward-propagating bow shock can be clearly seen as a sharp semi-elliptical discontinuity in the second and third rows of Figure \ref{plot3}, corresponding to the times $t= 250$, $300$ Myr, respectively. Along the $R$ direction, the shock's Mach number drops from $\sim 1.26$ at $t= 250$ Myr to $1.09$ at $t= 300$ Myr. Right behind the shock front, there is a shell of post-shock gas that is moving away from the cluster centre. This is essentially a rarefaction wave following the bow shock. The locations of the bow shock and rarefaction wave are also shown schematically in the middle two panels of Figure \ref{plot10} (see Sec. 3.6 for more details).

Here we stress important dynamical impacts of the AGN-driven bow shock and the post-shock rarefaction wave on the cool core and the development of cooling flows therein. As the bow shock propagates outward through the ICM, it reverses cooling inflows into outflows, and the shocked gas in the downstream of the shock forms the outward-propagating rarefaction wave, as clearly seen in the second and third rows of Figure \ref{plot3}. In our particular simulation at $t= 250$ or $300$ Myr, the typical outflowing velocity in the downstream of the shock is several hundreds km/s, which is much larger than the inflow speed of several tens km/s in the upstream of the shock. AGN-driven bow shocks have been frequently observed in cool-core clusters (\citealt{randall11}; \citealt{randall15}). The post-shock rarefaction wave causes cool core expansion and rarefies the dense cool core region, thus playing an important role in averting the central cooling catastrophe which otherwise develops in a classical cooling flow seen in our ``pure cooling" run.

As the shock front propagates far away, the shocked gas in inner regions changes its velocity direction again and moves inward, as seen in the fourth and fifth rows of Figure \ref{plot3}, corresponding to the times $t= 400$ and $500$ Myr, respectively. This is consistent with wave nature of the post-shock rarefaction wave. However, a particularly important kinetic feature within the post-shock cool core is the vertical outflow trailing behind the AGN bubble, as clearly seen in the bottom two rows of Figure \ref{plot3} (also see Fig. \ref{plot10}). These trailing outflows along the jet direction and inflows in other directions form cool-core circulations in meridional planes, which suppresses accumulation of gas in the cluster centre and the subsequent onset of the cooling catastrophe there. We will further investigate trailing outflows and meridional circulations in more detail in the next subsection. 

The AGN-driven expansion of cool cluster cores is real in our simulation, which can also be seen in Figure \ref{plot4}. The top panel shows the time evolution of normalized gas masses within three representative spheres of radii $r= 50$, $100$, and $200$ kpc. Before turning on the AGN jet, the total gas mass within each spherical region increases gradually with time, as a result of the cooling inflow. Once the jet is turned on at $t\sim 237.6$ Myr, the bow shock and rarefaction wave are produced, and the cluster expands from inner to outer regions. The gas mass enclosed within the inner $50$ kpc region decreases quickly by $\sim 30\%$ within several tens Myrs, while the total gas masses enclosed within $100$ and $200$ kpc regions start to decrease later as it takes time for the bow shock to propagate outward. The peak outflow rate is several thousands $M_{\odot}$/yr, as seen in the bottom panel, which also clearly shows that after the outflow episode, gas flows in again with negative values of $\dot{M}$. However, even at $t=500$ Myr, which is more than $250$ Myr after the onset of AGN feedback, the gas mass enclosed within the inner $50$ kpc is still less than its value at $t\sim 237.6$ Myr (right before the jet event), indicating that the overall cool core expansion induced by the jet event is significant.

\subsection{Energy Evolution and Shock Heating}

 \begin{figure}
\includegraphics[width=0.45\textwidth]{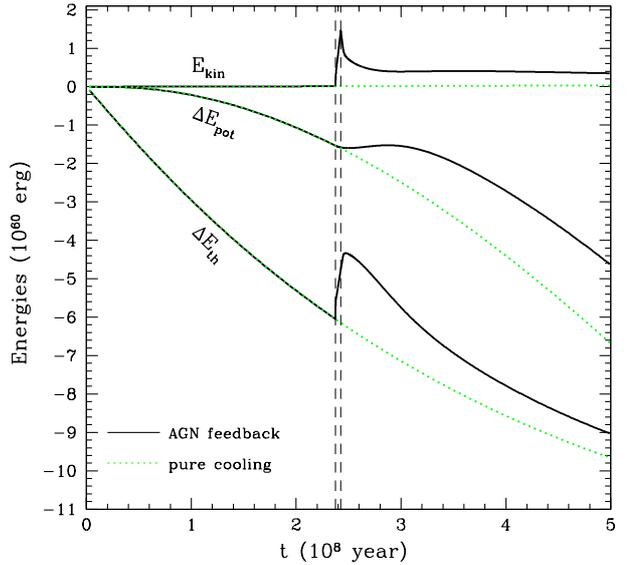} 
\caption{Global energy evolutions in the ``pure cooling" run ({\it dotted}) and our main ``AGN feedback" run ({\it solid}). From top to bottom, each solid (or dashed) line represents the temporal evolution of the kinetic energy $E_{\rm kin}$, the change in the potential energy $\Delta E_{\rm pot}$, and the change in thermal energy $\Delta E_{\rm th}$, all evaluated within the whole simulated hemisphere. The left and right vertical dashed lines indicate the start and end times of the AGN jet event, respectively.}
 \label{plot5}
 \end{figure} 

 \begin{figure}
\includegraphics[width=0.45\textwidth]{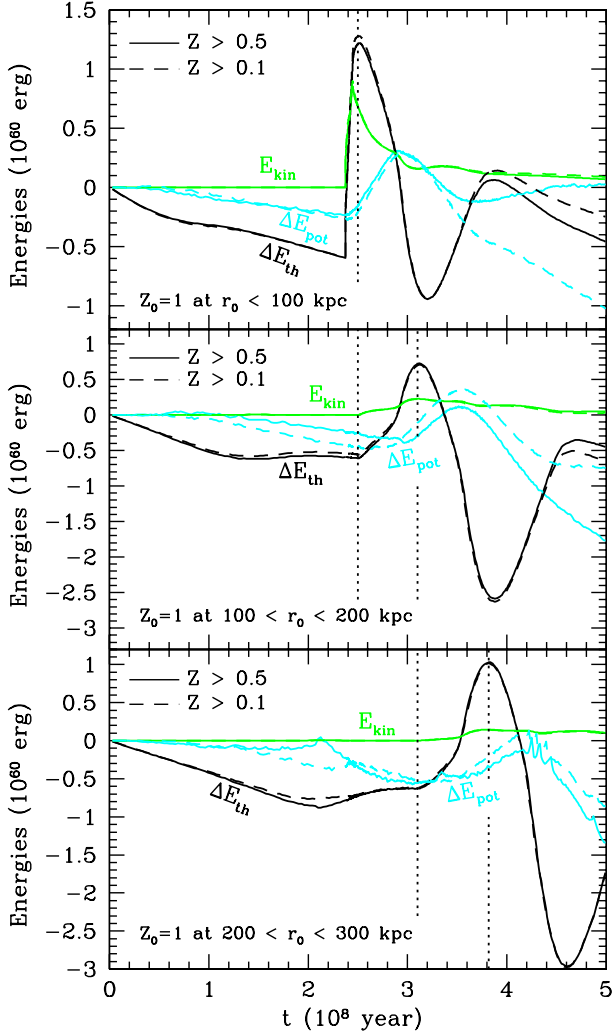} 
\caption{Temporal evolutions of the kinetic energy, change in the potential energy, and change in thermal energy, of the ICM gas elements with initial radii $r_{0}$ in the following three regions: $r_{0}<100$ kpc ({\it top panel}), $100<r_{0}<200$ kpc ({\it middle panel}), and $200<r_{0}<300$ kpc ({\it bottom panel}). For each panel, we performed a specific simulation where the gas elements in the corresponding region are traced in a Lagrangian way by a tracer fluid with initial concentrations $Z_{0}=1$ at $t=0$ ($Z_{0}$ is set to be zero for all other gas elements in the cluster). The value of $Z$ for each gas element is effectively conserved during evolution, although inevitable numerical diffusion tends to mix it with the ambient gas. Solid and dotted lines in each panel correspond to the temporal evolution of energies integrated from all the gas elements with $Z>0.5$ and $Z>0.1$, respectively, in the corresponding run.}
 \label{plot6}
 \end{figure} 

Here we study the energy evolution and heating mechanisms of the ICM during the gasdynamical processes investigated in the previous subsection. We first investigate the global energy evolution of the ICM in Figure \ref{plot5}, which shows the temporal evolution of the kinetic energy $E_{\rm kin}$, the change in the potential energy $\Delta E_{\rm pot}$, and the change in thermal energy $\Delta E_{\rm th}$ of the whole simulated ICM hemisphere in both the ``pure cooling" and AGN feedback runs. In the pure cooling run (dotted lines), the decrease of thermal energy is significant during the simulation, which is obviously due to radiative cooling. The cooling-induced inflow causes the cluster contraction, explaining the evident decrease of potential energy, which is gradually converted to thermal energy and kinetic energy. However, the increase of kinetic energy is negligibly small compared to changes in thermal and potential energies during the whole ``pure cooling" run.

In the AGN feedback run, an AGN jet is turned on at $t=237.6$ Myr (the left vertical dashed line), and lasts for $5$ Myr, causing sharp increases of the kinetic and thermal energies, as clearly seen in Figure \ref{plot5}. While the initial jet in the initialization nozzle is dominated by kinetic energy ($P_{\rm kin}/P_{\rm jet}\sim 97\%$), the increase of the system's thermal energy during the jet injection phase is comparable to that in kinetic energy, due to the rapid dissipation of kinetic energy by shocks within the supersonic jet and the bow shock. The kinetic energy of the system starts to decrease when the jet is turned off at $t=242.6$ Myr (the right vertical dashed line), while the increase of the thermal energy by shock dissipation lasts slightly longer until $t\sim245$ Myr. The potential energy initially decreases with time during the cooling inflow phase, but once the AGN jet is turned on, it becomes relatively flat with time until $t \sim 320$ Myr. The potential energy increases in the post-shock rarefaction wave, which is roughly compensated by its decrease in inflowing outer regions.

To further study jet heating on the cool core region, we show in the top panel of Figure \ref{plot6} the energy evolution of the ICM gas elements initially located with distances of $r_{0}<100$ kpc to the cluster centre. As described in detail in Section 3.5, we ``paint" these gas elements at $t=0$ with the only non-zero concentrations of a tracer fluid: $Z_{0}\equiv \rho'/\rho=1$, where $\rho'$ is the density of the tracer fluid. As each gas element moves, the value of concentration $Z$ is conserved, and thus we can study the energy evolution of these gas elements in a Lagrangian way. We note, however, that as the jet moves quickly across numerical grids in our simulation, it inevitably mixes with the ambient ICM gas through numerical diffusion (i.e., advection errors), and RT and KH instabilities at the bubble/ICM interface further facilitates the mixing, which results in some grids with $0<Z<1$ (see the top panel of Fig. \ref{plot8}). In the top panel of Figure \ref{plot6}, we show the temporal evolution of the energies associated with the gas elements with both $Z>0.5$ (solid) and $Z>0.1$ (dashed), and their evolutions are very similar (except for the late evolution of the potential energy), suggesting that our method does a reasonably good job in tracing the Lagrangian energy evolution of the ICM. 

Now let us look at the Lagrangian evolution of the kinetic energy, change in the potential energy, and change in thermal energy, of the gas elements with $r_{0}<100$ kpc in the top panel of Figure \ref{plot6}. When $t<237.6$ Myr, both the thermal and potential energies decrease with time due to cooling and the ICM contraction. Once the jet is turned on at $t=237.6$ Myr, both the kinetic and thermal energies increase rapidly. The increase phase of the kinetic energy is very brief, and stops at $t\sim245$ Myr, when shock dissipation surpasses kinetic motions induced by the jet injection. The vertical dotted line corresponds to $t=250$ Myr, when the bow shock starts to propagate across $r=100$ kpc (see the second row of Fig. \ref{plot3}) and the thermal energy starts to decrease due to adiabatic expansion in the rarefaction wave (also see Sec. 3.4). The peak thermal energy after the jet event is much higher than its initial value at $t=0$, indicating that the cool core is over-heated by the jet event. 

How does the jet heat the cool core gas in our simulation? Possible heating mechanisms include mixing, pdV work, turbulence, and shock. The jet-ICM mixing region is located within the low-density AGN bubble (cavity), and the ICM heating due to mixing should be appreciably less than the whole thermal energy contained in the AGN bubble (cavity). At $t=250$ Myr (the second row of Fig. \ref{plot3}), the AGN bubble may be approximated by a cylinder with radius $R_{b} \sim10$ kpc and height $h_{b}\sim 80$ kpc, and the internal thermal energy density $e_{b}$ varies between $2-5 \times 10^{-10}$ erg cm$^{-3}$. Taking $e_{b}\sim 4 \times 10^{-10}$ erg cm$^{-3}$, the whole thermal energy within the AGN bubble is $E_{b}\sim e_{b}\pi R_{b}^{2}h_{b}\sim 3\times 10^{59}$ erg, which is only about $1/6$ of the sharp increase of the ICM thermal energy $\Delta E_{\rm th} \sim 1.8\times 10^{60}$ erg from $t=237.6$ Myr to $250$ Myr. Similarly, the pdV work done during the inflation of the AGN bubble is $E_{\rm pdV} \sim P_{b}\pi R_{b}^{2}h_{b} \sim (\gamma-1)E_{b}\sim 2\times 10^{59}$ erg, which is only $\sim 10\%$ of $\Delta E_{\rm th}$. Here $P_{b}\sim (\gamma-1)e_{b}$ is the bubble pressure. We further argue that turbulence heating is unlikely to dominate (also see \citealt{reynolds15a}), as the velocity fields in Fig. \ref{plot3} do not show significant levels of turbulence and no explicit viscosity other than artificial viscosity is included to dissipate turbulence in our simulation. Furthermore, the dissipation of the jet's kinetic energy is very rapid. 

\begin{figure}
\includegraphics[width=0.45\textwidth]{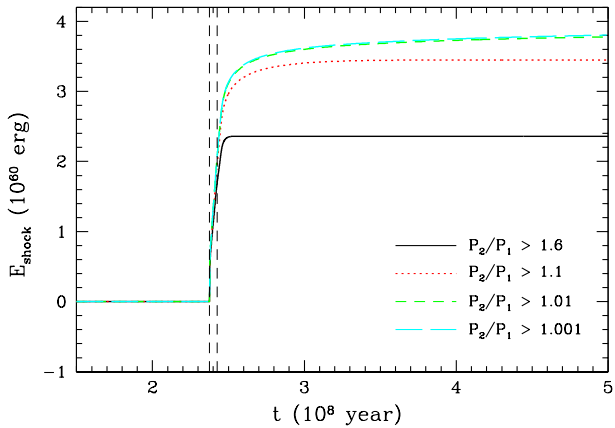} 
\caption{The total energy dissipated by shocks within time $t$ in all active zones with a pressure jump: $P_{\rm i}/P_{\rm i+1}>1.6$ (solid), $1.1$ (dotted), $1.01$ (short-dashed), and $1.001$ (long-dashed) along either the positive $z$ or $R$ direction. Shock heating is evaluated with the dissipation of kinetic energy by a von Neumann-Richtmyer artificial viscosity in the corresponding zones. The left and right vertical dashed lines indicate the start and end times of the AGN jet event, respectively.}
 \label{plot7}
 \end{figure}

We therefore conclude that the ICM heating in our simulation is mainly due to shock heating, which is consistent with the rather rapid dissipation of the jet's kinetic energy seen in Figure \ref{plot5} and the top panel of Figure \ref{plot6}. Figure \ref{plot7} shows the time evolution of the total energy $E_{\rm shock}$ dissipated by shocks within time $t$ in all active zones with a pressure jump: $P_{\rm i}/P_{\rm i+1}>1.6$ (solid), $1.1$ (dotted), $1.01$ (short-dashed), and $1.001$ (long-dashed), where $P_{\rm i}$ is the pressure in each zone and $P_{\rm i+1}$ is the pressure in the adjacent zone along either the positive $z$ or $R$ direction. As in the ZEUS code (\citealt{stone92}; also see \citealt{liyuan17}), shock dissipation is dealt with a von Neumann-Richtmyer artificial viscosity and $E_{\rm shock}$ can thus be simply evaluated with the dissipation of kinetic energy by the artificial viscosity in the corresponding zones. We note that $E_{\rm shock}$ includes shock dissipation within both the supersonic jet and ICM, although the reverse shock near the jet terminus ($P_{\rm i}/P_{\rm i+1}<1$) has already been excluded by the above pressure jump conditions. For very low values of $P_{\rm i}/P_{\rm i+1}$, $E_{\rm shock}$ may also include dissipation of complex flows and waves \citep{ryu03}. Nonetheless, Figure \ref{plot7} suggests that shock heating may indeed be significant, and happens mostly during and shortly after the jet event. For example, at $t=250$ Myr when the bow shock only reaches $R\sim 30$ kpc along the $R$ direction, shock dissipation with $P_{\rm i}/P_{\rm i+1}>1.6$ has almost finished, and $\sim 87\%$ of energy dissipation by shocks with $P_{\rm i}/P_{\rm i+1}>1.1$ has been completed. As clearly seen in the second and third rows of Figure \ref{plot3}, the most evident shock in the ICM is the bow shock, which presumably dominates over-heating of the cool core. Efficient shock heating during powerful AGN outbursts has also been found recently by \citet{yang16} and \citet{liyuan17}.

\subsection{Energy Transport by the Rarefaction Wave}

As studied in the previous subsection (Fig. \ref{plot5} and the top panel of Fig. \ref{plot6}), shocks driven by the powerful AGN jet dissipate kinetic energy rapidly and over-heat the cool cluster core in our simulation. This is consistent with earlier studies by \citet{fujita05} and \citet{mathews06}, who found that AGN-driven shocks tend to dissipate kinetic energy rapidly in inner cool core regions. Then, how does the jet transfer its energy to the whole cluster core and even outer cluster regions? Bow shocks are followed by rarefaction waves, which cause the cluster expansion. Here we study energy transport by the rarefaction wave in our simulation.

The top panel of Fig. \ref{plot6} shows that the thermal energy of the gas elements with $r_{0}<100$ kpc decreases significantly by an amount of $2.2\times 10^{60}$ erg from $t=0.25$ Gyr to $0.32$ Gyr, which is clearly due to adiabatic losses by the core expansion during the rarefaction wave phase. Presumably a significant fraction of the lost energy is transported to outer regions. To confirm this, we follow the energy evolution of the gas elements initially located within $100<r_{0}<200$ kpc at $t=0$ in the middle panel of Fig. \ref{plot6}, which clearly shows that the thermal energy of these gas elements increases significantly from $t=0.25$ Gyr (the left vertical dotted line) to $0.31$ Gyr (the right vertical dotted line) by an amount of $1.3 \times 10^{60}$ erg. In the rarefaction wave, some thermal energy is also converted to potential energy, while another fraction is lost to radiation.

At $t\sim 0.31$ Gyr, the shock front reaches $r \sim 200$ kpc, and the thermal energy of the gas elements with $100<r_{0}<200$ kpc decreases significantly from $t=0.31$ Gyr to $0.38$ Gyr by an amount of $3.3 \times 10^{60}$ erg due to the gas expansion. Most of the lost energy is transported to the gas elements with $r_{0}<100$ kpc (see the corresponding thermal energy increase during the same epoch in the top panel) and $200<r_{0}<300$ kpc (see the thermal energy increase between two dotted lines in the bottom panel of Fig. \ref{plot6}). At $t\sim 0.38$ Gyr, the thermal energy of the cool core gas with $r_{0}<100$ kpc is close to its initial value at $t=0$, indicating that the net energy gain during the first $380$ Myr through shock heating, adiabatic expansion, and the later contraction roughly balances the energy loss through radiation. As also seen in the bottom panel of Fig. \ref{plot6}, the bow shock and rarefaction wave reach $r \sim 300$ kpc at $t\sim 0.38$ Gyr, and the thermal energy of the gas elements with $200<r_{0}<300$ kpc starts to decrease significantly as it is adiabatically transported to the inner ($100<r_{0}<200$ kpc) and outer ($r_{0}>300$ kpc) regions. We note that, to make the middle and bottom panels, we performed two additional simulations that trace the Lagrangian evolution of the gas elements initially located in the regions $100<r_{0}<200$ kpc and $200<r_{0}<300$ kpc, respectively, by assigning the only nonzero values of $Z_{0}=1$ to each corresponding region.

In summary, the rarefaction wave which follows the bow shock transports thermal energy from the over-heated cluster core to outer regions, and thus plays a key role in the energy redistribution in galaxy clusters during AGN feedback events. 

\subsection{Trailing Outflows and Meridional Circulations}

 \begin{figure}
\includegraphics[width=0.45\textwidth]{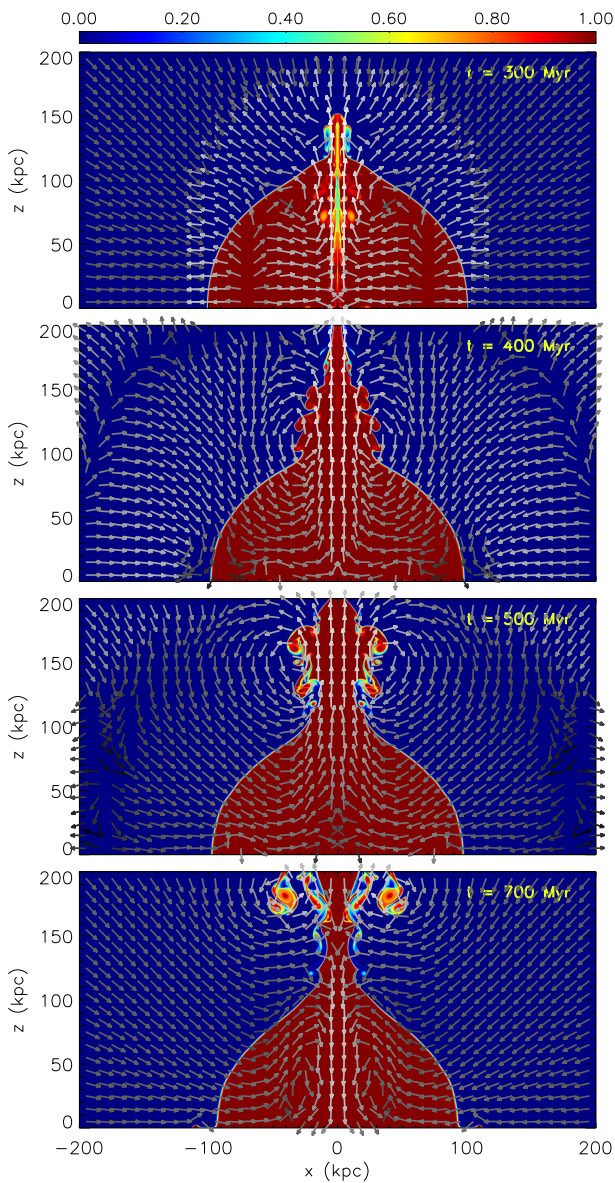} 
\caption{Temporal evolution of the concentration distribution ($Z$) of the tracer fluid in a simulation where the initial value of $Z$ at $t=0$ is $1$ if $r\leq100$ kpc and $0$ if $r >100$ kpc. Arrows over-plotted in each panel indicate directions of gas velocities. From top to bottom, the time of each panel is $t=300$, $400$, $500$, $700$ Myr. The top colour bar represents $Z$ in linear scale. It is clear that trailing flows behind the AGN bubble form real outflows in the ICM, which fall back to the cluster centre at later times, e.g., at $t=700$ Myr.}
 \label{plot8}
 \end{figure} 

In this subsection, we study in more detail {\it trailing outflows} of AGN bubbles and {\it meridional circulations} in cool cluster cores found in Section 3.2. The ICM velocity distributions in the bottom two rows of Figure \ref{plot3} clearly show coherent gas outflows behind the AGN bubble along the $z$ axis at $t=400$ and $500$ Myr, extending from the cluster centre to the bottom of the AGN bubble ($\sim 150$ kpc in our run). In the present paper, we refer these coherent outflows as ``intracluster trailing outflows", as they are apparently trailing flows of AGN bubbles in their wakes. We note that in our simulation the AGN bubble is significantly disrupted by Rayleigh-Taylor (RT) and Kelvin-Helmholtz (KH) instabilities \citep{reynolds05}, unlike some old intact X-ray cavities detected in galaxy clusters (e.g., the outer northwestern X-ray cavity in Perseus; see Figure 3 in \citealt{fabian00}). The suppression of these interface instabilities during the development of AGN bubbles may be due to shear viscosity (\citealt{kaiser05}; \citealt{reynolds05}; \citealt{guo12b}; \citealt{guo15}) or magnetic tension (\citealt{kaiser05}; \citealt{jones05}; \citealt{ruszkowski07}), which is not included in our simulation. However, the formation of trailing outflows should be robust, as also seen in simulations of AGN bubbles in viscous media (see Fig. 8 of \citealt{guo15}).

Do coherent outflowing motions trailing behind the AGN bubble represent real outflows? To answer this question, we rerun our simulation by additionally following the evolution of a tracer fluid $\rho'$ which follows gas motions, i.e., ${d \rho'}/{d t} + \rho' \nabla \cdot {\bf v} = 0$. Combining with the gas mass conservation equation (\ref{hydro1}), one can easily derive the governing equation for the concentration of the tracer fluid ($Z\equiv \rho'/\rho$): $dZ/dt=0$, i.e., the value of $Z$ for each fluid element does not change with time. In our simulation, we choose as the initial condition at $t=0$ the values of $Z$ to be:
\begin{eqnarray}
Z_{0}=
\begin{cases}
1 & \text{  if } r\leq100\text{ kpc,}
\\
0 & \text{  if } r>100 \text{ kpc.}
\end{cases}
\end{eqnarray}
Thus the concentration $Z$ can be used to trace the spatial evolution of the ICM gas initially located within the central sphere of radius $100$ kpc. 

Figure \ref{plot8} shows the time evolution of the concentration distribution ($Z$) of the tracer fluid. It is clear that the trailing outflow is real: some of the gas originally located within $r\leq100$ kpc is uplifted to $r\sim100 -  200$ kpc along the jet direction by the AGN bubble at $t=400$, $500$, $700$ Myr. The trailing outflow along the $z$ direction and inflows in other directions form gas circulations in the meridional planes, and near the bottom of the AGN bubble, meridional vortexes also appear. At $t=700$ Myr, one can clearly see that the trailing flow is falling back to the cluster centre (see the overplotted velocity vectors), suggesting that the outflow does not leave the cluster eventually, but instead it is confined to the cluster.

Trailing outflows play an important role in suppressing or delaying the central cooling catastrophe. As the rarefaction wave propagates away to large radii, gas inflows resume in the inner core region, as clearly seen in Figures \ref{plot3} and \ref{plot8} and discussed in Section 3.2. If there are no trailing outflows, the gas density in the cluster centre would rise rapidly and a central cooling catastrophe would then develop. Trailing outflows uplifted by AGN bubbles transport gas from the inner cluster core to outer regions, suppressing gas accumulation at the cluster centre resulting from inflows along other directions and thus averting the central cooling catastrophe, which is delayed until trailing outflows fall back to the cluster centre at later times (see Section 3.6).

Observational evidence for trailing outflows has been found in some cool core clusters where the spatial distributions of heavy elements are anisotropic and preferentially aligned with the large-scale radio and cavity axes (\citealt{kirkpatrick11}; \citealt{kirkpatrick15}), suggesting that hot metal-rich outflows are indeed lifted up by AGN jets. Hydrodynamic simulations following both the jet and metallicity evolution in galaxy clusters would be helpful to further explore this scenario (e.g., \citealt{barai16}, which however does not investigate the angular distribution of metals). Along the axes of trailing outflows, gas density is relatively high, as seen in the bottom two rows of Figure \ref{plot3}, and cold gas may drop out due to local thermal instability, forming cold filaments found in some galaxy clusters (e.g., \citealt{mittal12}; \citealt{tremblay15}).

 \begin{figure}
\includegraphics[width=0.45\textwidth]{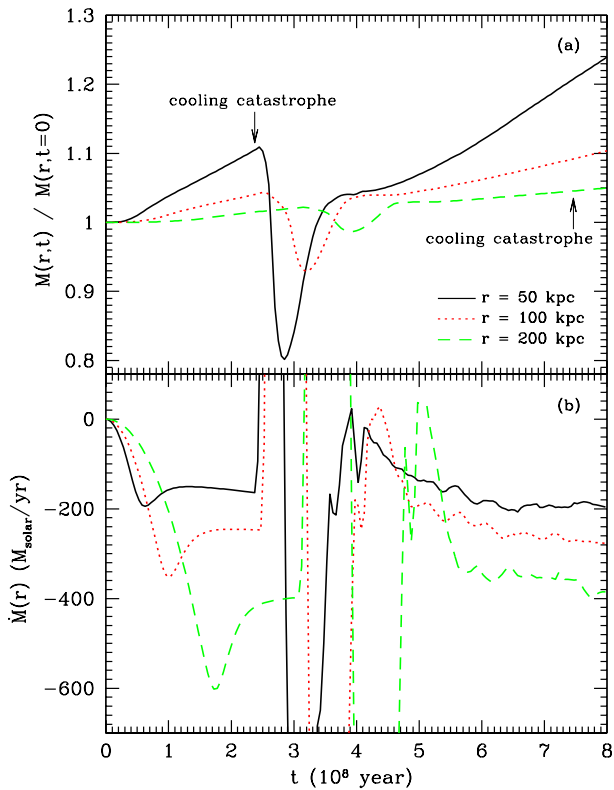} 
\caption{Same as Figure \ref{plot4}, but showing evolution to a later time $t=800$ Myr. The two arrows in the top panel indicate the onset times of the first and second cooling catastrophes. An AGN jet event is manually turned on right after the first cooling catastrophe lasting for a duration of $t_{\rm jet}=5$ Myr, while no AGN jets are turned on after the second cooling catastrophe. The bottom panel shows that the mass inflow rate across $r=50$ kpc in the developed cooling flow of A1795 is about $160$ - $200 M_{\odot}/$yr. 
}
 \label{plot9}
 \end{figure}

\subsection{Emerging Picture of Cyclic Cool Cores}

 \begin{figure*}
\includegraphics[trim=0 0 0 0.5, clip,width=0.9\textwidth]{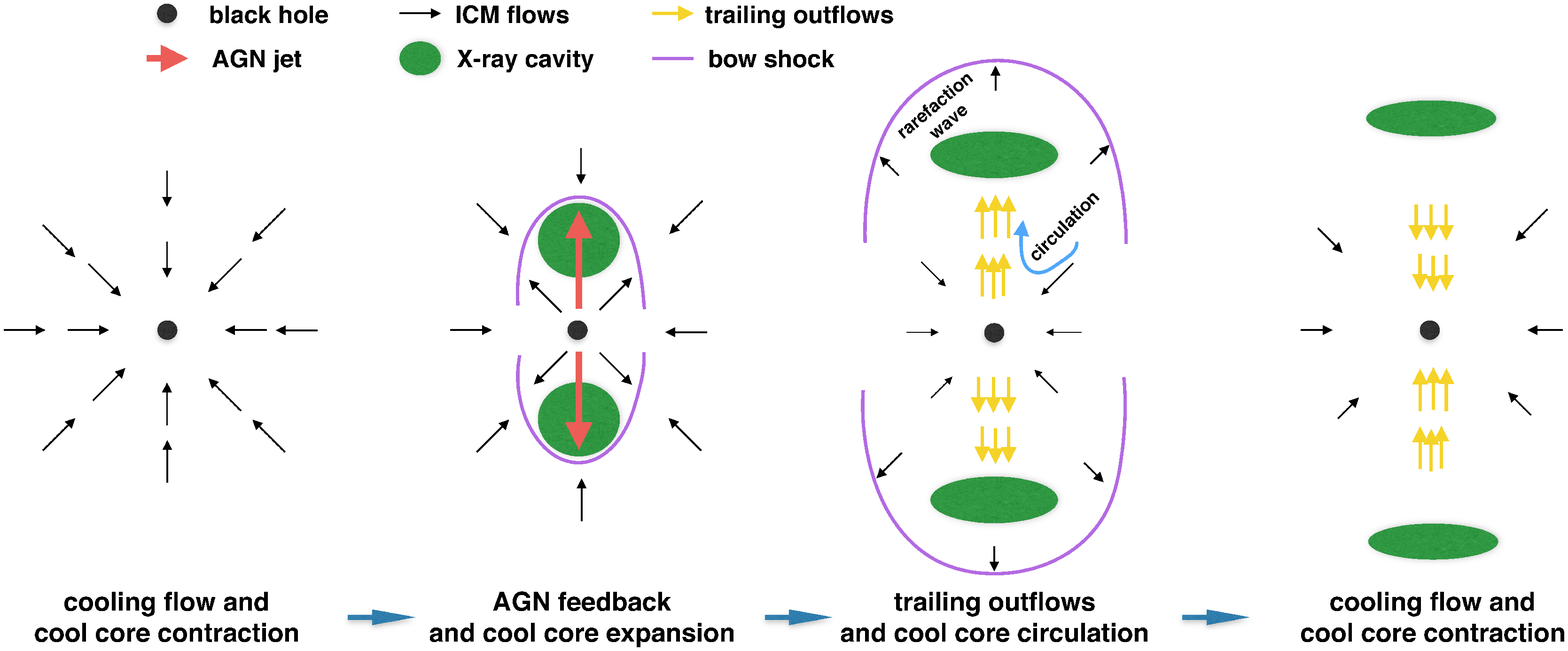} 
\caption{Schematic picture of the cyclic cool core of a galaxy cluster, regulated by cooling flows and AGN feedback. Cooling inflows lead to cool core contraction and the central cooling catastrophe, which triggers a pair of AGN jets, resulting in AGN bubbles and bow shocks. The outward bow shock reverses cooling inflows into outflows, and over-heats the cool core. The following rarefaction wave leads to cool core expansion and adiabatically transports energy to outer regions. The rising AGN bubbles uplift trailing outflows behind them, forming meridional gas circulations in the cluster core, as shown in the third panel. At later times, trailing outflows fall back toward the cluster centre, and together with cooling inflows, the cluster core contracts again, leading to another central cooling catastrophe, which may trigger a new-generation AGN feedback event. }
 \label{plot10}
 \end{figure*}
 
What is the long-term evolution of cool cluster cores after powerful AGN feedback events? After the onset of the AGN jet event, the cool core in our simulation is quickly over-heated by shock dissipation, and rarefied by the rarefaction wave which transports energy to large radii. As the bow shock weakens and propagates further away, the inner core region starts to contract again and form meridional circulations with trailing outflows uplifted by the AGN bubble, as shown in Figure \ref{plot3}. Trailing outflows avoid rapid accumulation of gas at the cluster centre, and thus delay the onset of central cooling catastrophe. However, as shown by the velocity field in the bottom panel of Figure \ref{plot8}, the trailing outflows could not eventually leave the deep potential well of the cluster, and at $t=700$ Myr, they are clearly falling back to the cluster centre along the $z$ axis. The gas density at the cluster centre would then increase rapidly, and at $t=746$ Myr in our simulation, a cooling catastrophe happens again in the cluster centre. The temporal evolution of gas masses within three representative spheres and the mass flow rates across their surfaces are shown in Figure \ref{plot9}, indicating that the mass inflow rate across $r=50$ kpc in the developed cooling flow in our simulated cluster A1795 is about $160$ - $200 M_{\odot}/$yr. The mass inflow rate in the cooling flow phase increases with radius, suggesting that gas densities in the cool core increase continuously, as also seen in the left panel of Figure \ref{plot2}. 
 
Thus, the evolution of cool cluster cores seems to be cyclic, and a typical cycle is summarized schematically in Figure \ref{plot10}. What would be observational evidence for cyclic cool cores going through cooling-induced contraction and AGN-induced expansion. Cool cores are under-heated during the cooling flow phase, but over-heated during the AGN outburst phase. Thus in the former phase, gas densities and the X-ray luminosity of the cool core are relatively higher, while the central gas entropy would be relatively lower. For galaxy clusters with the same virial mass, cyclic cool cores suggested by our simulation would thus lead to an anti-correlation between the gas fraction (or X-ray luminosity) of cool cores and the central gas entropy, which is not naturally expected in steady-state models of cool cluster cores.

\section{Summary and Discussion}
\label{section:discussion}

We investigate the cooling flow problem in a typical cool core cluster -- Abell 1795, and in particular, focus on how a powerful AGN jet event triggered by the central cooling catastrophe, heats and reverses the cluster core from the ``catastrophic cooling" state back to its initial cool-core state. We use a ``pure cooling" control run where no AGN jets are triggered for comparison, and investigate in detail complex gasdynamical processes and energy evolution in an ``AGN feedback" run, where a powerful AGN jet event is turned on once the central cooling catastrophe starts. The jet parameters are specifically chosen so that the cluster profiles will reverse back to their initial distributions after the jet event.The supersonic jet is active steadily for a duration of $5$ Myr, and its power $P_{\rm jet}=1.5\times10^{46}$ erg/s, which is dominated by the kinetic energy, is comparable to that estimated in Cygnus A \citep{wilson06} and MS0735 \citep{mcnamara09}, but less than the maximum jet power $\sim 10^{47}$ erg/s reached in some recent simulations (e.g., \citealt{gaspari11}; \citealt{li15}; \citealt{yang16b}). Our main findings are summarized below:

\begin{itemize}

 \item{Without AGN feedback, a cooling inflow gradually develops, and cooling catastrophe first appears at the cluster centre (at $t\sim 237.6$ Myr in our simulation), which is consistent with previous studies (e.g., \citealt{li12} and \citealt{guo14}).}
 
 \item{The AGN jet triggered after the central cooling catastrophe drives a bow shock, which reverses cooling inflows into outflowing motions. A rarefaction wave following the shock front expands the shocked ICM, rarefying the dense cool core and lowering central cooling rates.}

\item{The powerful AGN jet event is able to bring the cluster core from the catastrophic cooling state back to the initial cool-core state. In our simulation, ICM heating is dominated by shock heating, while mixing and turbulent heating are less important. Most shock energy is dissipated rapidly in the inner core region, while the rarefaction wave is efficient in transporting and re-distributing energy to outer regions.}

 \item{As the AGN bubble rises away from the cluster centre, it uplifts trailing outflows in its wake along the original jet direction. Since inflows resume in the core along other directions as the rarefaction wave propagates away, trailing outflows suppress gas accumulation and cooling rates in the cluster centre by transporting gas from the inner core to outer regions. As a result, meridional circulations also form in the cool core.}
 
 \item{Trailing outflows are intracluster (i.e., confined to galaxy clusters), and at later times, a fraction of them fall back to the cluster centre, triggering central cooling catastrophe and potentially a new generation of AGN feedback.} 

\item{We propose a picture of cyclic cool cores that cycle through cooling-induced contraction and AGN-induced expansion. During the former state, cool cores are under-heated, while during the latter, they are over-heated. This picture thus naturally leads to an anti-correlation between the gas fraction (or X-ray luminosity) of cool cores and the central gas entropy, which may be tested by X-ray observations. }

\end{itemize}

We discuss several limitations of our work here. First, our simulations focus on a specific powerful AGN jet event, while in reality AGN feedback events have a variety of powers and durations. The mechanical variability of AGN outbursts is also quite uncertain. If an AGN jet episode is less powerful but lasts for a longer duration, AGN heating and dynamical feedback would be more gentle. In this case, the dominant heating mechanism and the following energy redistribution may be different than in the present paper, which we leave for future studies.

For simplicity, the parameters of the jet in its initialization nozzle are fixed with time in our simulation when the jet is active. While this assumption potentially corresponds to a steady accretion flow onto the central SMBH after cold gas circularizes around it, it does not reflect a long-term feedback mechanism that automatically links gas cooling, SMBH accretion, and AGN outbursts. Therefore, although our simulation indicates that a powerful AGN jet event is able to reverse a galaxy cluster from the catastrophic cooaling state back to the initial cool-core state, we could not yet answer if cooling flows can be averted by AGN feedback over a long, multi-Gyr timescale. Recent simulations tackling this long-term cooling flow problem often invoke jet precession to avoid the formation of a low-density channel along the jet direction that could rapidly transport jet energy out of cool cores  (\citealt{gaspari12}; \citealt{li15}; \citealt{yang16b}), while dynamic ICM weather in realistic clusters may also help distribute jet energy to broader angular regions (\citealt{mendygral11}; \citealt{mendygral12}).

In our simulation, gas motions induced by the AGN jet event are quite regular, and the level of turbulence is not significant, which is consistent with recent studies by \citet{reynolds15a}, \citet{yang16b} and \citet{liyuan17}. However, we note that simulations with grid-based codes inevitably introduce numerical diffusion, which may suppress the development of turbulence. A careful investigation on this effect is required to fully understand the importance of turbulence in AGN feedback.

Finally, our simulation only includes pure hydrodynamics, but neglects some more subtle physics, e.g., viscosity, magnetic fields, cosmic rays, etc. Cosmic rays and magnetic fields are present and may even dominate in AGN jets and bubbles (\citealt{guo08a}; \citealt{guo11}; \citealt{ruszkowski17}; \citealt{gan17}). Shear viscosity (\citealt{kaiser05}; \citealt{reynolds05}; \citealt{guo12b}; \citealt{guo15}) and magnetic tension (\citealt{kaiser05}; \citealt{jones05}; \citealt{ruszkowski07}) tend to prevent the disruption of AGN bubbles by suppressing the development of RT and KH instabilities, which may indeed happen in some old intact X-ray cavities detected in galaxy clusters. Nonetheless, we expect that our main results on complex gasdynamical processes induced by AGN jet events, including shocks, rarefaction waves, trailing outflows, and meridional circulations, and their importance in solving the cooling flow problem should be robust.

\section*{Acknowledgments}

FG thanks James Binney, Fabrizio Brighenti, and Eugene Churazov for helpful discussions. We thank the referee for a useful report that helped to improve the manuscript. This work was supported by Chinese Academy of Sciences through the Hundred Talents Program and the Frontier Sciences Key Project (No. QYZDB-SSW-SYS033), and National Natural Science Foundation of China (Grant No. 11643001 and 11633006). YFY gracefully acknowledges support from National Natural Science Foundation of China (Grant No. U1431228, 11133005, 11233003, and 11421303). Some simulations presented in this work were performed using the high performance computing resources in the Core Facility for Advanced Research Computing at Shanghai Astronomical Observatory.

\bibliography{ms}

\begin{thebibliography}{}
\makeatletter
\relax
\def\mn@urlcharsother{\let\do\@makeother \do\$\do\&\do\#\do\^\do\_\do\%\do\~}
\def\mn@doi{\begingroup\mn@urlcharsother \@ifnextchar [ {\mn@doi@}
  {\mn@doi@[]}}
\def\mn@doi@[#1]#2{\def\@tempa{#1}\ifx\@tempa\@empty \href
  {http://dx.doi.org/#2} {doi:#2}\else \href {http://dx.doi.org/#2} {#1}\fi
  \endgroup}
\def\mn@eprint#1#2{\mn@eprint@#1:#2::\@nil}
\def\mn@eprint@arXiv#1{\href {http://arxiv.org/abs/#1} {{\tt arXiv:#1}}}
\def\mn@eprint@dblp#1{\href {http://dblp.uni-trier.de/rec/bibtex/#1.xml}
  {dblp:#1}}
\def\mn@eprint@#1:#2:#3:#4\@nil{\def\@tempa {#1}\def\@tempb {#2}\def\@tempc
  {#3}\ifx \@tempc \@empty \let \@tempc \@tempb \let \@tempb \@tempa \fi \ifx
  \@tempb \@empty \def\@tempb {arXiv}\fi \@ifundefined
  {mn@eprint@\@tempb}{\@tempb:\@tempc}{\expandafter \expandafter \csname
  mn@eprint@\@tempb\endcsname \expandafter{\@tempc}}}

\bibitem[\protect\citeauthoryear{{Anninos}, {Zhang}, {Abel}  \&
  {Norman}}{{Anninos} et~al.}{1997}]{anninos97}
{Anninos} P.,  {Zhang} Y.,  {Abel} T.,   {Norman} M.~L.,  1997, \mn@doi [\na]
  {10.1016/S1384-1076(97)00009-2}, \href
  {http://adsabs.harvard.edu/abs/1997NewA....2..209A} {2, 209}

\bibitem[\protect\citeauthoryear{{Babul}, {Sharma}  \& {Reynolds}}{{Babul}
  et~al.}{2013}]{babul13}
{Babul} A.,  {Sharma} P.,   {Reynolds} C.~S.,  2013, \mn@doi [\apj]
  {10.1088/0004-637X/768/1/11}, \href
  {http://adsabs.harvard.edu/abs/2013ApJ...768...11B} {768, 11}

\bibitem[\protect\citeauthoryear{{Barai}, {Murante}, {Borgani}, {Gaspari},
  {Granato}, {Monaco}  \& {Ragone-Figueroa}}{{Barai} et~al.}{2016}]{barai16}
{Barai} P.,  {Murante} G.,  {Borgani} S.,  {Gaspari} M.,  {Granato} G.~L.,
  {Monaco} P.,   {Ragone-Figueroa} C.,  2016, \mn@doi [\mnras]
  {10.1093/mnras/stw1389}, \href
  {http://adsabs.harvard.edu/abs/2016MNRAS.461.1548B} {461, 1548}

\bibitem[\protect\citeauthoryear{{B{\^i}rzan}, {Rafferty}, {McNamara}, {Wise}
  \& {Nulsen}}{{B{\^i}rzan} et~al.}{2004}]{birzan04}
{B{\^i}rzan} L.,  {Rafferty} D.~A.,  {McNamara} B.~R.,  {Wise} M.~W.,
  {Nulsen} P.~E.~J.,  2004, \mn@doi [\apj] {10.1086/383519}, \href
  {http://adsabs.harvard.edu/abs/2004ApJ...607..800B} {607, 800}

\bibitem[\protect\citeauthoryear{{Boehringer}, {Voges}, {Fabian}, {Edge}  \&
  {Neumann}}{{Boehringer} et~al.}{1993}]{boehringer93}
{Boehringer} H.,  {Voges} W.,  {Fabian} A.~C.,  {Edge} A.~C.,   {Neumann}
  D.~M.,  1993, \mnras, \href
  {http://adsabs.harvard.edu/abs/1993MNRAS.264L..25B} {264, L25}

\bibitem[\protect\citeauthoryear{{Brighenti} \& {Mathews}}{{Brighenti} \&
  {Mathews}}{2006}]{brighenti06}
{Brighenti} F.,  {Mathews} W.~G.,  2006, \mn@doi [\apj] {10.1086/502645}, \href
  {http://adsabs.harvard.edu/abs/2006ApJ...643..120B} {643, 120}

\bibitem[\protect\citeauthoryear{{Br{\"u}ggen}, {Heinz}, {Roediger},
  {Ruszkowski}  \& {Simionescu}}{{Br{\"u}ggen} et~al.}{2007a}]{bruggen07}
{Br{\"u}ggen} M.,  {Heinz} S.,  {Roediger} E.,  {Ruszkowski} M.,   {Simionescu}
  A.,  2007a, \mn@doi [\mnras] {10.1111/j.1745-3933.2007.00351.x}, \href
  {http://adsabs.harvard.edu/abs/2007MNRAS.380L..67B} {380, L67}

\bibitem[\protect\citeauthoryear{{Br{\"u}ggen}, {Heinz}, {Roediger},
  {Ruszkowski}  \& {Simionescu}}{{Br{\"u}ggen} et~al.}{2007b}]{brueggen07}
{Br{\"u}ggen} M.,  {Heinz} S.,  {Roediger} E.,  {Ruszkowski} M.,   {Simionescu}
  A.,  2007b, \mn@doi [\mnras] {10.1111/j.1745-3933.2007.00351.x}, \href
  {http://adsabs.harvard.edu/abs/2007MNRAS.380L..67B} {380, L67}

\bibitem[\protect\citeauthoryear{{Cattaneo} \& {Teyssier}}{{Cattaneo} \&
  {Teyssier}}{2007}]{cattaneo07}
{Cattaneo} A.,  {Teyssier} R.,  2007, \mn@doi [\mnras]
  {10.1111/j.1365-2966.2007.11512.x}, \href
  {http://adsabs.harvard.edu/abs/2007MNRAS.376.1547C} {376, 1547}

\bibitem[\protect\citeauthoryear{{Chandran} \& {Rasera}}{{Chandran} \&
  {Rasera}}{2007}]{chandran07}
{Chandran} B.~D.~G.,  {Rasera} Y.,  2007, \mn@doi [\apj] {10.1086/521619},
  \href {http://adsabs.harvard.edu/abs/2007ApJ...671.1413C} {671, 1413}

\bibitem[\protect\citeauthoryear{{Croston}, {Hardcastle}, {Mingo}, {Evans},
  {Dicken}, {Morganti}  \& {Tadhunter}}{{Croston} et~al.}{2011}]{croston11}
{Croston} J.~H.,  {Hardcastle} M.~J.,  {Mingo} B.,  {Evans} D.~A.,  {Dicken}
  D.,  {Morganti} R.,   {Tadhunter} C.~N.,  2011, \mn@doi [\apjl]
  {10.1088/2041-8205/734/2/L28}, \href
  {http://adsabs.harvard.edu/abs/2011ApJ...734L..28C} {734, L28}

\bibitem[\protect\citeauthoryear{{Dubois}, {Devriendt}, {Teyssier}  \&
  {Slyz}}{{Dubois} et~al.}{2011}]{dubois11}
{Dubois} Y.,  {Devriendt} J.,  {Teyssier} R.,   {Slyz} A.,  2011, \mn@doi
  [\mnras] {10.1111/j.1365-2966.2011.19381.x}, \href
  {http://adsabs.harvard.edu/abs/2011MNRAS.417.1853D} {417, 1853}

\bibitem[\protect\citeauthoryear{{Ettori}, {Fabian}, {Allen}  \&
  {Johnstone}}{{Ettori} et~al.}{2002}]{ettori02}
{Ettori} S.,  {Fabian} A.~C.,  {Allen} S.~W.,   {Johnstone} R.~M.,  2002,
  \mn@doi [\mnras] {10.1046/j.1365-8711.2002.05212.x}, \href
  {http://adsabs.harvard.edu/abs/2002MNRAS.331..635E} {331, 635}

\bibitem[\protect\citeauthoryear{{Fabian}}{{Fabian}}{2012}]{fabian12}
{Fabian} A.~C.,  2012, \mn@doi [\araa] {10.1146/annurev-astro-081811-125521},
  \href {http://adsabs.harvard.edu/abs/2012ARA%26A..50..455F} {50, 455}

\bibitem[\protect\citeauthoryear{{Fabian} et~al.,}{{Fabian}
  et~al.}{2000}]{fabian00}
{Fabian} A.~C.,  et~al., 2000, \mn@doi [\mnras]
  {10.1046/j.1365-8711.2000.03904.x}, \href
  {http://adsabs.harvard.edu/abs/2000MNRAS.318L..65F} {318, L65}

\bibitem[\protect\citeauthoryear{{Fabian}, {Celotti}, {Blundell}, {Kassim}  \&
  {Perley}}{{Fabian} et~al.}{2002}]{fabian02}
{Fabian} A.~C.,  {Celotti} A.,  {Blundell} K.~M.,  {Kassim} N.~E.,   {Perley}
  R.~A.,  2002, \mn@doi [\mnras] {10.1046/j.1365-8711.2002.05182.x}, \href
  {http://adsabs.harvard.edu/abs/2002MNRAS.331..369F} {331, 369}

\bibitem[\protect\citeauthoryear{{Fabian}, {Sanders}, {Allen}, {Crawford},
  {Iwasawa}, {Johnstone}, {Schmidt}  \& {Taylor}}{{Fabian}
  et~al.}{2003}]{fabian03b}
{Fabian} A.~C.,  {Sanders} J.~S.,  {Allen} S.~W.,  {Crawford} C.~S.,  {Iwasawa}
  K.,  {Johnstone} R.~M.,  {Schmidt} R.~W.,   {Taylor} G.~B.,  2003, \mn@doi
  [\mnras] {10.1046/j.1365-8711.2003.06902.x}, \href
  {http://adsabs.harvard.edu/abs/2003MNRAS.344L..43F} {344, L43}

\bibitem[\protect\citeauthoryear{{Fabian}, {Walker}, {Russell}, {Pinto},
  {Sanders}  \& {Reynolds}}{{Fabian} et~al.}{2017}]{fabian17}
{Fabian} A.~C.,  {Walker} S.~A.,  {Russell} H.~R.,  {Pinto} C.,  {Sanders}
  J.~S.,   {Reynolds} C.~S.,  2017, \mn@doi [\mnras] {10.1093/mnrasl/slw170},
  \href {http://adsabs.harvard.edu/abs/2017MNRAS.464L...1F} {464, L1}

\bibitem[\protect\citeauthoryear{{Fujita} \& {Suzuki}}{{Fujita} \&
  {Suzuki}}{2005}]{fujita05}
{Fujita} Y.,  {Suzuki} T.~K.,  2005, \mn@doi [\apjl] {10.1086/491649}, \href
  {http://adsabs.harvard.edu/abs/2005ApJ...630L...1F} {630, L1}

\bibitem[\protect\citeauthoryear{{Gan}, {Li}, {Li}  \& {Yuan}}{{Gan}
  et~al.}{2017}]{gan17}
{Gan} Z.,  {Li} H.,  {Li} S.,   {Yuan} F.,  2017, \mn@doi [\apj]
  {10.3847/1538-4357/aa647e}, \href
  {http://adsabs.harvard.edu/abs/2017ApJ...839...14G} {839, 14}

\bibitem[\protect\citeauthoryear{{Gaspari} \& {S{\c a}dowski}}{{Gaspari} \&
  {S{\c a}dowski}}{2017}]{gaspari17}
{Gaspari} M.,  {S{\c a}dowski} A.,  2017, \mn@doi [\apj]
  {10.3847/1538-4357/aa61a3}, \href
  {http://adsabs.harvard.edu/abs/2017ApJ...837..149G} {837, 149}

\bibitem[\protect\citeauthoryear{{Gaspari}, {Melioli}, {Brighenti}  \&
  {D'Ercole}}{{Gaspari} et~al.}{2011}]{gaspari11}
{Gaspari} M.,  {Melioli} C.,  {Brighenti} F.,   {D'Ercole} A.,  2011, \mn@doi
  [\mnras] {10.1111/j.1365-2966.2010.17688.x}, \href
  {http://adsabs.harvard.edu/abs/2011MNRAS.411..349G} {411, 349}

\bibitem[\protect\citeauthoryear{{Gaspari}, {Ruszkowski}  \&
  {Sharma}}{{Gaspari} et~al.}{2012}]{gaspari12}
{Gaspari} M.,  {Ruszkowski} M.,   {Sharma} P.,  2012, \mn@doi [\apj]
  {10.1088/0004-637X/746/1/94}, \href
  {http://adsabs.harvard.edu/abs/2012ApJ...746...94G} {746, 94}

\bibitem[\protect\citeauthoryear{{Gitti}, {McNamara}, {Nulsen}  \&
  {Wise}}{{Gitti} et~al.}{2007}]{gitti07}
{Gitti} M.,  {McNamara} B.~R.,  {Nulsen} P.~E.~J.,   {Wise} M.~W.,  2007,
  \mn@doi [\apj] {10.1086/512800}, \href
  {http://adsabs.harvard.edu/abs/2007ApJ...660.1118G} {660, 1118}

\bibitem[\protect\citeauthoryear{{Guo}}{{Guo}}{2015}]{guo15}
{Guo} F.,  2015, \mn@doi [\apj] {10.1088/0004-637X/803/1/48}, \href
  {http://adsabs.harvard.edu/abs/2015ApJ...803...48G} {803, 48}

\bibitem[\protect\citeauthoryear{{Guo}}{{Guo}}{2016}]{guo16}
{Guo} F.,  2016, \mn@doi [\apj] {10.3847/0004-637X/826/1/17}, \href
  {http://adsabs.harvard.edu/abs/2016ApJ...826...17G} {826, 17}

\bibitem[\protect\citeauthoryear{{Guo} \& {Mathews}}{{Guo} \&
  {Mathews}}{2010a}]{guo10a}
{Guo} F.,  {Mathews} W.~G.,  2010a, \mn@doi [\apj]
  {10.1088/0004-637X/712/2/1311}, \href
  {http://adsabs.harvard.edu/abs/2010ApJ...712.1311G} {712, 1311}

\bibitem[\protect\citeauthoryear{{Guo} \& {Mathews}}{{Guo} \&
  {Mathews}}{2010b}]{guo10b}
{Guo} F.,  {Mathews} W.~G.,  2010b, \mn@doi [\apj]
  {10.1088/0004-637X/717/2/937}, \href
  {http://adsabs.harvard.edu/abs/2010ApJ...717..937G} {717, 937}

\bibitem[\protect\citeauthoryear{{Guo} \& {Mathews}}{{Guo} \&
  {Mathews}}{2011}]{guo11}
{Guo} F.,  {Mathews} W.~G.,  2011, \mn@doi [\apj]
  {10.1088/0004-637X/728/2/121}, \href
  {http://adsabs.harvard.edu/abs/2011ApJ...728..121G} {728, 121}

\bibitem[\protect\citeauthoryear{{Guo} \& {Mathews}}{{Guo} \&
  {Mathews}}{2012}]{guo12}
{Guo} F.,  {Mathews} W.~G.,  2012, \mn@doi [\apj]
  {10.1088/0004-637X/756/2/181}, \href
  {http://adsabs.harvard.edu/abs/2012ApJ...756..181G} {756, 181}

\bibitem[\protect\citeauthoryear{{Guo} \& {Mathews}}{{Guo} \&
  {Mathews}}{2014}]{guo14}
{Guo} F.,  {Mathews} W.~G.,  2014, \mn@doi [\apj]
  {10.1088/0004-637X/780/2/126}, \href
  {http://adsabs.harvard.edu/abs/2014ApJ...780..126G} {780, 126}

\bibitem[\protect\citeauthoryear{{Guo} \& {Oh}}{{Guo} \& {Oh}}{2008}]{guo08a}
{Guo} F.,  {Oh} S.~P.,  2008, \mn@doi [\mnras]
  {10.1111/j.1365-2966.2007.12692.x}, \href
  {http://adsabs.harvard.edu/abs/2008MNRAS.384..251G} {384, 251}

\bibitem[\protect\citeauthoryear{{Guo}, {Oh}  \& {Ruszkowski}}{{Guo}
  et~al.}{2008}]{guo08b}
{Guo} F.,  {Oh} S.~P.,   {Ruszkowski} M.,  2008, \mn@doi [\apj]
  {10.1086/592320}, \href {http://adsabs.harvard.edu/abs/2008ApJ...688..859G}
  {688, 859}

\bibitem[\protect\citeauthoryear{{Guo}, {Mathews}, {Dobler}  \& {Oh}}{{Guo}
  et~al.}{2012}]{guo12b}
{Guo} F.,  {Mathews} W.~G.,  {Dobler} G.,   {Oh} S.~P.,  2012, \mn@doi [\apj]
  {10.1088/0004-637X/756/2/182}, \href
  {http://adsabs.harvard.edu/abs/2012ApJ...756..182G} {756, 182}

\bibitem[\protect\citeauthoryear{{Hardcastle} \& {Krause}}{{Hardcastle} \&
  {Krause}}{2013}]{hardcastle13}
{Hardcastle} M.~J.,  {Krause} M.~G.~H.,  2013, \mn@doi [\mnras]
  {10.1093/mnras/sts564}, \href
  {http://adsabs.harvard.edu/abs/2013MNRAS.430..174H} {430, 174}

\bibitem[\protect\citeauthoryear{{Heckman} \& {Best}}{{Heckman} \&
  {Best}}{2014}]{heckman14}
{Heckman} T.~M.,  {Best} P.~N.,  2014, \mn@doi [\araa]
  {10.1146/annurev-astro-081913-035722}, \href
  {http://adsabs.harvard.edu/abs/2014ARA%26A..52..589H} {52, 589}

\bibitem[\protect\citeauthoryear{{Hernquist}}{{Hernquist}}{1990}]{hernquist90}
{Hernquist} L.,  1990, \mn@doi [\apj] {10.1086/168845}, \href
  {http://adsabs.harvard.edu/abs/1990ApJ...356..359H} {356, 359}

\bibitem[\protect\citeauthoryear{{Hillel} \& {Soker}}{{Hillel} \&
  {Soker}}{2016}]{hillel16}
{Hillel} S.,  {Soker} N.,  2016, \mn@doi [\mnras] {10.1093/mnras/stv2483},
  \href {http://adsabs.harvard.edu/abs/2016MNRAS.455.2139H} {455, 2139}

\bibitem[\protect\citeauthoryear{{Hillel} \& {Soker}}{{Hillel} \&
  {Soker}}{2017}]{hillel17}
{Hillel} S.,  {Soker} N.,  2017, \mn@doi [\mnras] {10.1093/mnrasl/slw231},
  \href {http://adsabs.harvard.edu/abs/2017MNRAS.466L..39H} {466, L39}

\bibitem[\protect\citeauthoryear{{Jacob} \& {Pfrommer}}{{Jacob} \&
  {Pfrommer}}{2017}]{jacob17}
{Jacob} S.,  {Pfrommer} C.,  2017, \mn@doi [\mnras] {10.1093/mnras/stx131},
  \href {http://adsabs.harvard.edu/abs/2017MNRAS.467.1449J} {467, 1449}

\bibitem[\protect\citeauthoryear{{Jones} \& {De Young}}{{Jones} \& {De
  Young}}{2005}]{jones05}
{Jones} T.~W.,  {De Young} D.~S.,  2005, \mn@doi [\apj] {10.1086/429157}, \href
  {http://adsabs.harvard.edu/abs/2005ApJ...624..586J} {624, 586}

\bibitem[\protect\citeauthoryear{{Kaiser}, {Pavlovski}, {Pope}  \&
  {Fangohr}}{{Kaiser} et~al.}{2005}]{kaiser05}
{Kaiser} C.~R.,  {Pavlovski} G.,  {Pope} E.~C.~D.,   {Fangohr} H.,  2005,
  \mn@doi [\mnras] {10.1111/j.1365-2966.2005.08902.x}, \href
  {http://adsabs.harvard.edu/abs/2005MNRAS.359..493K} {359, 493}

\bibitem[\protect\citeauthoryear{{Kirkpatrick} \& {McNamara}}{{Kirkpatrick} \&
  {McNamara}}{2015}]{kirkpatrick15}
{Kirkpatrick} C.~C.,  {McNamara} B.~R.,  2015, \mn@doi [\mnras]
  {10.1093/mnras/stv1574}, \href
  {http://adsabs.harvard.edu/abs/2015MNRAS.452.4361K} {452, 4361}

\bibitem[\protect\citeauthoryear{{Kirkpatrick}, {McNamara}  \&
  {Cavagnolo}}{{Kirkpatrick} et~al.}{2011}]{kirkpatrick11}
{Kirkpatrick} C.~C.,  {McNamara} B.~R.,   {Cavagnolo} K.~W.,  2011, \mn@doi
  [\apjl] {10.1088/2041-8205/731/2/L23}, \href
  {http://adsabs.harvard.edu/abs/2011ApJ...731L..23K} {731, L23}

\bibitem[\protect\citeauthoryear{{Li} \& {Bryan}}{{Li} \& {Bryan}}{2012}]{li12}
{Li} Y.,  {Bryan} G.~L.,  2012, \mn@doi [\apj] {10.1088/0004-637X/747/1/26},
  \href {http://adsabs.harvard.edu/abs/2012ApJ...747...26L} {747, 26}

\bibitem[\protect\citeauthoryear{{Li} \& {Bryan}}{{Li} \& {Bryan}}{2014}]{li14}
{Li} Y.,  {Bryan} G.~L.,  2014, \mn@doi [\apj] {10.1088/0004-637X/789/1/54},
  \href {http://adsabs.harvard.edu/abs/2014ApJ...789...54L} {789, 54}

\bibitem[\protect\citeauthoryear{{Li}, {Bryan}, {Ruszkowski}, {Voit}, {O'Shea}
  \& {Donahue}}{{Li} et~al.}{2015}]{li15}
{Li} Y.,  {Bryan} G.~L.,  {Ruszkowski} M.,  {Voit} G.~M.,  {O'Shea} B.~W.,
  {Donahue} M.,  2015, \mn@doi [\apj] {10.1088/0004-637X/811/2/73}, \href
  {http://adsabs.harvard.edu/abs/2015ApJ...811...73L} {811, 73}

\bibitem[\protect\citeauthoryear{{Li}, {Ruszkowski}  \& {Bryan}}{{Li}
  et~al.}{2016}]{liyuan17}
{Li} Y.,  {Ruszkowski} M.,   {Bryan} G.~L.,  2016, ArXiv: 1611.05455, \href
  {http://adsabs.harvard.edu/abs/2016arXiv161105455L} {}

\bibitem[\protect\citeauthoryear{{Mathews} \& {Guo}}{{Mathews} \&
  {Guo}}{2010}]{mathews10}
{Mathews} W.~G.,  {Guo} F.,  2010, \mn@doi [\apj]
  {10.1088/0004-637X/725/2/1440}, \href
  {http://adsabs.harvard.edu/abs/2010ApJ...725.1440M} {725, 1440}

\bibitem[\protect\citeauthoryear{{Mathews}, {Faltenbacher}  \&
  {Brighenti}}{{Mathews} et~al.}{2006}]{mathews06}
{Mathews} W.~G.,  {Faltenbacher} A.,   {Brighenti} F.,  2006, \mn@doi [\apj]
  {10.1086/499119}, \href {http://adsabs.harvard.edu/abs/2006ApJ...638..659M}
  {638, 659}

\bibitem[\protect\citeauthoryear{{McNamara} \& {Nulsen}}{{McNamara} \&
  {Nulsen}}{2012}]{mcnamara12}
{McNamara} B.~R.,  {Nulsen} P.~E.~J.,  2012, \mn@doi [New Journal of Physics]
  {10.1088/1367-2630/14/5/055023}, \href
  {http://cdsads.u-strasbg.fr/abs/2012NJPh...14e5023M} {14, 055023}

\bibitem[\protect\citeauthoryear{{McNamara}, {Kazemzadeh}, {Rafferty},
  {B{\^i}rzan}, {Nulsen}, {Kirkpatrick}  \& {Wise}}{{McNamara}
  et~al.}{2009}]{mcnamara09}
{McNamara} B.~R.,  {Kazemzadeh} F.,  {Rafferty} D.~A.,  {B{\^i}rzan} L.,
  {Nulsen} P.~E.~J.,  {Kirkpatrick} C.~C.,   {Wise} M.~W.,  2009, \mn@doi
  [\apj] {10.1088/0004-637X/698/1/594}, \href
  {http://adsabs.harvard.edu/abs/2009ApJ...698..594M} {698, 594}

\bibitem[\protect\citeauthoryear{{Mendygral}, {O'Neill}  \&
  {Jones}}{{Mendygral} et~al.}{2011}]{mendygral11}
{Mendygral} P.~J.,  {O'Neill} S.~M.,   {Jones} T.~W.,  2011, \mn@doi [\apj]
  {10.1088/0004-637X/730/2/100}, \href
  {http://adsabs.harvard.edu/abs/2011ApJ...730..100M} {730, 100}

\bibitem[\protect\citeauthoryear{{Mendygral}, {Jones}  \& {Dolag}}{{Mendygral}
  et~al.}{2012}]{mendygral12}
{Mendygral} P.~J.,  {Jones} T.~W.,   {Dolag} K.,  2012, \mn@doi [\apj]
  {10.1088/0004-637X/750/2/166}, \href
  {http://adsabs.harvard.edu/abs/2012ApJ...750..166M} {750, 166}

\bibitem[\protect\citeauthoryear{{Mittal} et~al.,}{{Mittal}
  et~al.}{2012}]{mittal12}
{Mittal} R.,  et~al., 2012, \mn@doi [\mnras]
  {10.1111/j.1365-2966.2012.21891.x}, \href
  {http://adsabs.harvard.edu/abs/2012MNRAS.426.2957M} {426, 2957}

\bibitem[\protect\citeauthoryear{{Navarro}, {Frenk}  \& {White}}{{Navarro}
  et~al.}{1997}]{navarro97}
{Navarro} J.~F.,  {Frenk} C.~S.,   {White} S.~D.~M.,  1997, \mn@doi [\apj]
  {10.1086/304888}, \href {http://adsabs.harvard.edu/abs/1997ApJ...490..493N}
  {490, 493}

\bibitem[\protect\citeauthoryear{{Norman}, {Winkler}, {Smarr}  \&
  {Smith}}{{Norman} et~al.}{1982}]{norman82}
{Norman} M.~L.,  {Winkler} K.-H.~A.,  {Smarr} L.,   {Smith} M.~D.,  1982, \aap,
  \href {http://adsabs.harvard.edu/abs/1982A%26A...113..285N} {113, 285}

\bibitem[\protect\citeauthoryear{{Paczy{\'n}sky} \& {Wiita}}{{Paczy{\'n}sky} \&
  {Wiita}}{1980}]{paczynsky80}
{Paczy{\'n}sky} B.,  {Wiita} P.~J.,  1980, \aap, \href
  {http://adsabs.harvard.edu/abs/1980A%26A....88...23P} {88, 23}

\bibitem[\protect\citeauthoryear{{Peterson} \& {Fabian}}{{Peterson} \&
  {Fabian}}{2006}]{peterson06}
{Peterson} J.~R.,  {Fabian} A.~C.,  2006, \mn@doi [\physrep]
  {10.1016/j.physrep.2005.12.007}, \href
  {http://adsabs.harvard.edu/cgi-bin/nph-bib_query?bibcode=2006PhR...427....1P&db_key=AST}
  {427, 1}

\bibitem[\protect\citeauthoryear{{Prasad}, {Sharma}  \& {Babul}}{{Prasad}
  et~al.}{2015}]{prasad15}
{Prasad} D.,  {Sharma} P.,   {Babul} A.,  2015, \mn@doi [\apj]
  {10.1088/0004-637X/811/2/108}, \href
  {http://adsabs.harvard.edu/abs/2015ApJ...811..108P} {811, 108}

\bibitem[\protect\citeauthoryear{{Proga}, {MacFadyen}, {Armitage}  \&
  {Begelman}}{{Proga} et~al.}{2003}]{proga03}
{Proga} D.,  {MacFadyen} A.~I.,  {Armitage} P.~J.,   {Begelman} M.~C.,  2003,
  \mn@doi [\apjl] {10.1086/381158}, \href
  {http://adsabs.harvard.edu/abs/2003ApJ...599L...5P} {599, L5}

\bibitem[\protect\citeauthoryear{{Quataert} \& {Narayan}}{{Quataert} \&
  {Narayan}}{2000}]{quataert00}
{Quataert} E.,  {Narayan} R.,  2000, \mn@doi [\apj] {10.1086/308171}, \href
  {http://adsabs.harvard.edu/abs/2000ApJ...528..236Q} {528, 236}

\bibitem[\protect\citeauthoryear{{Rafferty}, {McNamara}, {Nulsen}  \&
  {Wise}}{{Rafferty} et~al.}{2006}]{rafferty06}
{Rafferty} D.~A.,  {McNamara} B.~R.,  {Nulsen} P.~E.~J.,   {Wise} M.~W.,  2006,
  \mn@doi [\apj] {10.1086/507672}, \href
  {http://adsabs.harvard.edu/abs/2006ApJ...652..216R} {652, 216}

\bibitem[\protect\citeauthoryear{{Randall} et~al.,}{{Randall}
  et~al.}{2011}]{randall11}
{Randall} S.~W.,  et~al., 2011, \mn@doi [\apj] {10.1088/0004-637X/726/2/86},
  \href {http://adsabs.harvard.edu/abs/2011ApJ...726...86R} {726, 86}

\bibitem[\protect\citeauthoryear{{Randall} et~al.,}{{Randall}
  et~al.}{2015}]{randall15}
{Randall} S.~W.,  et~al., 2015, \mn@doi [\apj] {10.1088/0004-637X/805/2/112},
  \href {http://adsabs.harvard.edu/abs/2015ApJ...805..112R} {805, 112}

\bibitem[\protect\citeauthoryear{{Reynolds}, {Heinz}  \& {Begelman}}{{Reynolds}
  et~al.}{2001}]{reynolds01}
{Reynolds} C.~S.,  {Heinz} S.,   {Begelman} M.~C.,  2001, \mn@doi [\apjl]
  {10.1086/319159}, \href {http://adsabs.harvard.edu/abs/2001ApJ...549L.179R}
  {549, L179}

\bibitem[\protect\citeauthoryear{{Reynolds}, {McKernan}, {Fabian}, {Stone}  \&
  {Vernaleo}}{{Reynolds} et~al.}{2005}]{reynolds05}
{Reynolds} C.~S.,  {McKernan} B.,  {Fabian} A.~C.,  {Stone} J.~M.,   {Vernaleo}
  J.~C.,  2005, \mn@doi [\mnras] {10.1111/j.1365-2966.2005.08643.x}, \href
  {http://adsabs.harvard.edu/abs/2005MNRAS.357..242R} {357, 242}

\bibitem[\protect\citeauthoryear{{Reynolds}, {Balbus}  \&
  {Schekochihin}}{{Reynolds} et~al.}{2015}]{reynolds15a}
{Reynolds} C.~S.,  {Balbus} S.~A.,   {Schekochihin} A.~A.,  2015, \mn@doi
  [\apj] {10.1088/0004-637X/815/1/41}, \href
  {http://adsabs.harvard.edu/abs/2015ApJ...815...41R} {815, 41}

\bibitem[\protect\citeauthoryear{{Ruszkowski} \& {Begelman}}{{Ruszkowski} \&
  {Begelman}}{2002}]{ruszkowski02}
{Ruszkowski} M.,  {Begelman} M.~C.,  2002, \mn@doi [\apj] {10.1086/344170},
  \href {http://adsabs.harvard.edu/abs/2002ApJ...581..223R} {581, 223}

\bibitem[\protect\citeauthoryear{{Ruszkowski}, {Br{\"u}ggen}  \&
  {Begelman}}{{Ruszkowski} et~al.}{2004}]{ruszkowski04a}
{Ruszkowski} M.,  {Br{\"u}ggen} M.,   {Begelman} M.~C.,  2004, \mn@doi [\apj]
  {10.1086/422158}, \href
  {http://adsabs.harvard.edu/cgi-bin/nph-bib_query?bibcode=2004ApJ...611..158R&db_key=AST}
  {611, 158}

\bibitem[\protect\citeauthoryear{{Ruszkowski}, {En{\ss}lin}, {Br{\"u}ggen},
  {Heinz}  \& {Pfrommer}}{{Ruszkowski} et~al.}{2007}]{ruszkowski07}
{Ruszkowski} M.,  {En{\ss}lin} T.~A.,  {Br{\"u}ggen} M.,  {Heinz} S.,
  {Pfrommer} C.,  2007, \mn@doi [\mnras] {10.1111/j.1365-2966.2007.11801.x},
  \href {http://adsabs.harvard.edu/abs/2007MNRAS.378..662R} {378, 662}

\bibitem[\protect\citeauthoryear{{Ruszkowski}, {Yang}  \&
  {Reynolds}}{{Ruszkowski} et~al.}{2017}]{ruszkowski17}
{Ruszkowski} M.,  {Yang} H.-Y.~K.,   {Reynolds} C.~S.,  2017, ArXiv:
  1701.07441, \href {http://adsabs.harvard.edu/abs/2017arXiv170107441R} {}

\bibitem[\protect\citeauthoryear{{Ryu}, {Kang}, {Hallman}  \& {Jones}}{{Ryu}
  et~al.}{2003}]{ryu03}
{Ryu} D.,  {Kang} H.,  {Hallman} E.,   {Jones} T.~W.,  2003, \mn@doi [\apj]
  {10.1086/376723}, \href {http://adsabs.harvard.edu/abs/2003ApJ...593..599R}
  {593, 599}

\bibitem[\protect\citeauthoryear{{Soker}}{{Soker}}{2016}]{soker16}
{Soker} N.,  2016, \mn@doi [\nar] {10.1016/j.newar.2016.08.002}, \href
  {http://adsabs.harvard.edu/abs/2016NewAR..75....1S} {75, 1}

\bibitem[\protect\citeauthoryear{{Stone} \& {Norman}}{{Stone} \&
  {Norman}}{1992}]{stone92}
{Stone} J.~M.,  {Norman} M.~L.,  1992, \mn@doi [\apjs] {10.1086/191680}, \href
  {http://adsabs.harvard.edu/cgi-bin/nph-bib_query?bibcode=1992ApJS...80..753S&db_key=AST}
  {80, 753}

\bibitem[\protect\citeauthoryear{{Sutherland} \& {Dopita}}{{Sutherland} \&
  {Dopita}}{1993}]{sd93}
{Sutherland} R.~S.,  {Dopita} M.~A.,  1993, \mn@doi [\apjs] {10.1086/191823},
  \href
  {http://adsabs.harvard.edu/cgi-bin/nph-bib_query?bibcode=1993ApJS...88..253S&db_key=AST}
  {88, 253}

\bibitem[\protect\citeauthoryear{{Tamura} et~al.,}{{Tamura}
  et~al.}{2001}]{tamura01}
{Tamura} T.,  et~al., 2001, \mn@doi [\aap] {10.1051/0004-6361:20000038}, \href
  {http://adsabs.harvard.edu/abs/2001A%26A...365L..87T} {365, L87}

\bibitem[\protect\citeauthoryear{{Tang} \& {Churazov}}{{Tang} \&
  {Churazov}}{2017}]{tang17}
{Tang} X.,  {Churazov} E.,  2017, \mn@doi [\mnras] {10.1093/mnras/stx590},
  \href {http://adsabs.harvard.edu/abs/2017MNRAS.468.3516T} {468, 3516}

\bibitem[\protect\citeauthoryear{{Tremblay} et~al.,}{{Tremblay}
  et~al.}{2015}]{tremblay15}
{Tremblay} G.~R.,  et~al., 2015, \mn@doi [\mnras] {10.1093/mnras/stv1151},
  \href {http://adsabs.harvard.edu/abs/2015MNRAS.451.3768T} {451, 3768}

\bibitem[\protect\citeauthoryear{{Vernaleo} \& {Reynolds}}{{Vernaleo} \&
  {Reynolds}}{2006}]{vernaleo06}
{Vernaleo} J.~C.,  {Reynolds} C.~S.,  2006, \mn@doi [\apj] {10.1086/504029},
  \href {http://adsabs.harvard.edu/abs/2006ApJ...645...83V} {645, 83}

\bibitem[\protect\citeauthoryear{{Vikhlinin}, {Kravtsov}, {Forman}, {Jones},
  {Markevitch}, {Murray}  \& {Van Speybroeck}}{{Vikhlinin}
  et~al.}{2006}]{vikhlinin06}
{Vikhlinin} A.,  {Kravtsov} A.,  {Forman} W.,  {Jones} C.,  {Markevitch} M.,
  {Murray} S.~S.,   {Van Speybroeck} L.,  2006, \mn@doi [\apj]
  {10.1086/500288}, \href {http://adsabs.harvard.edu/abs/2006ApJ...640..691V}
  {640, 691}

\bibitem[\protect\citeauthoryear{{Wilson}, {Smith}  \& {Young}}{{Wilson}
  et~al.}{2006}]{wilson06}
{Wilson} A.~S.,  {Smith} D.~A.,   {Young} A.~J.,  2006, \mn@doi [\apjl]
  {10.1086/504108}, \href {http://adsabs.harvard.edu/abs/2006ApJ...644L...9W}
  {644, L9}

\bibitem[\protect\citeauthoryear{{Yang} \& {Reynolds}}{{Yang} \&
  {Reynolds}}{2016a}]{yang16}
{Yang} H.-Y.~K.,  {Reynolds} C.~S.,  2016a, \mn@doi [\apj]
  {10.3847/0004-637X/818/2/181}, \href
  {http://adsabs.harvard.edu/abs/2016ApJ...818..181Y} {818, 181}

\bibitem[\protect\citeauthoryear{{Yang} \& {Reynolds}}{{Yang} \&
  {Reynolds}}{2016b}]{yang16b}
{Yang} H.-Y.~K.,  {Reynolds} C.~S.,  2016b, \mn@doi [\apj]
  {10.3847/0004-637X/829/2/90}, \href
  {http://adsabs.harvard.edu/abs/2016ApJ...829...90Y} {829, 90}

\bibitem[\protect\citeauthoryear{{Yuan} \& {Narayan}}{{Yuan} \&
  {Narayan}}{2014}]{yuan14}
{Yuan} F.,  {Narayan} R.,  2014, \mn@doi [\araa]
  {10.1146/annurev-astro-082812-141003}, \href
  {http://adsabs.harvard.edu/abs/2014ARA%26A..52..529Y} {52, 529}

\bibitem[\protect\citeauthoryear{{Zhuravleva} et~al.,}{{Zhuravleva}
  et~al.}{2014}]{zhuravleva14}
{Zhuravleva} I.,  et~al., 2014, \mn@doi [\nat] {10.1038/nature13830}, \href
  {http://adsabs.harvard.edu/abs/2014Natur.515...85Z} {515, 85}

\bibitem[\protect\citeauthoryear{{Zhuravleva} et~al.,}{{Zhuravleva}
  et~al.}{2016}]{zhuravleva16}
{Zhuravleva} I.,  et~al., 2016, \mn@doi [\mnras] {10.1093/mnras/stw520}, \href
  {http://adsabs.harvard.edu/abs/2016MNRAS.458.2902Z} {458, 2902}

\makeatother
\end{thebibliography}

\label{lastpage}

\end{document}